\documentclass{ieeetmlcn}

\usepackage{cite}
\usepackage{amsmath,amssymb,amsfonts}
\usepackage{algorithmic}
\usepackage{graphicx,color}
\usepackage{textcomp}
\usepackage{xcolor}
\usepackage{hyperref}
\hypersetup{hidelinks=true}

\usepackage{savesym}
\savesymbol{labelindent}
\usepackage{enumitem}
\restoresymbol{enumi}{labelindent}

\usepackage{multirow}
\usepackage{tabularx}

\definecolor{NavyBlue}{rgb}{0.0, 0.0, 0.5}

\usepackage{amssymb} 
\usepackage{array}
\usepackage{makecell}

\usepackage{color, colortbl}
\definecolor{aliceblue}{rgb}{0.94, 0.97, 1.0}
\definecolor{paleaqua}{rgb}{0.74, 0.83, 0.9}
\definecolor{palecyan}{rgb}{0.6, 0.8, 0.85}

\usepackage{balance}

\usepackage[ruled,linesnumbered]{algorithm2e}

\SetAlCapNameFnt{\scriptsize}
\SetAlCapFnt{\scriptsize}

\SetCommentSty{mycommfont}

\def\BibTeX{{\rm B\kern-.05em{\sc i\kern-.025em b}\kern-.08em
    T\kern-.1667em\lower.7ex\hbox{E}\kern-.125emX}}
    
\def\BibTeX{{\rm B\kern-.05em{\sc i\kern-.025em b}\kern-.08em T\kern-.1667em\lower.7ex\hbox{E}\kern-.125emX}}

\newcolumntype{M}[1]{>{\centering\arraybackslash}m{#1}}

\def\OJlogo{\vspace{-4pt}\includegraphics[height=18pt]{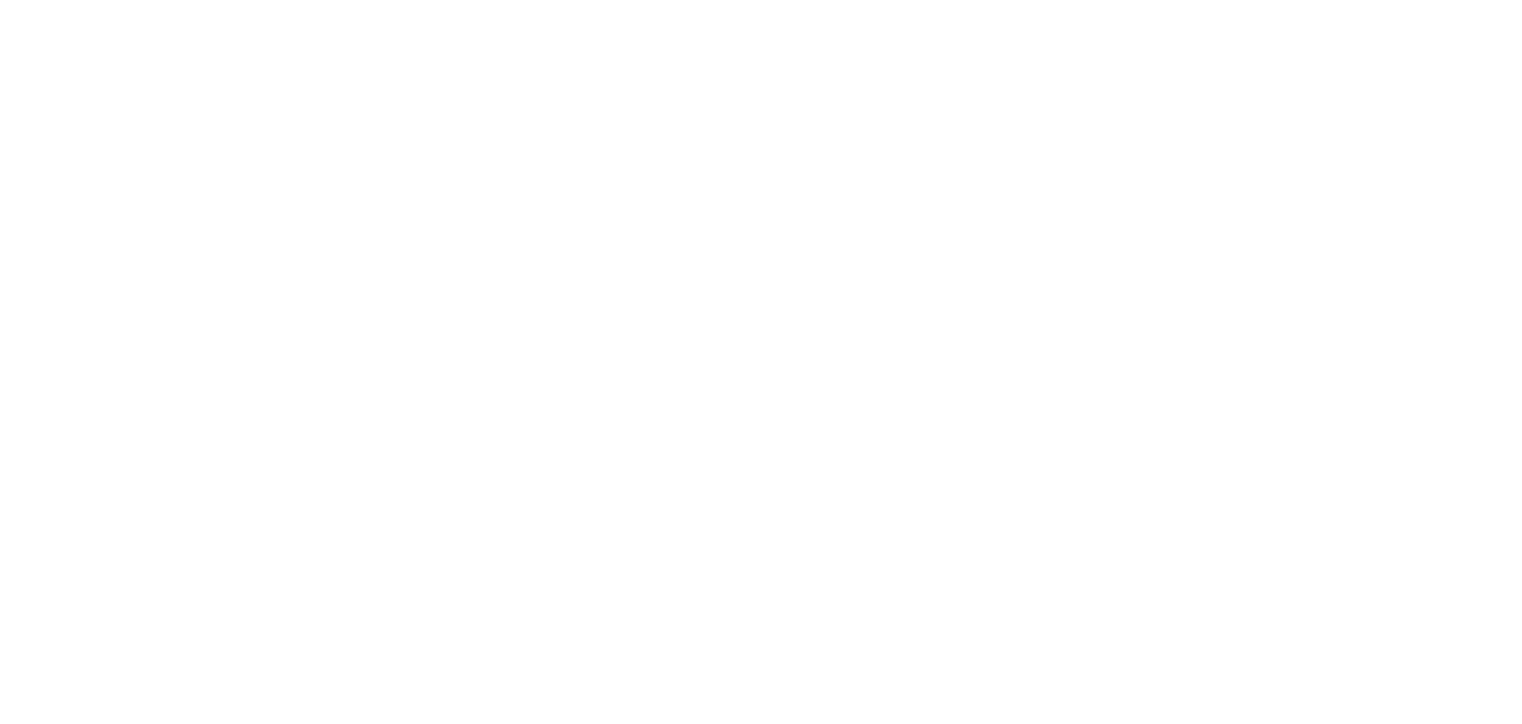}}
\def\seclogo{\vspace{10pt}\includegraphics[height=18pt]{new_logo}}

\def\authorrefmark#1{\ensuremath{^{\textbf{#1}}}}

\begin{document}

\markboth{}{Li Yang and Abdallah Shami}

\title{Toward Autonomous and Efficient Cybersecurity: A Multi-Objective AutoML-based Intrusion Detection System}

\author{ \textbf{L}i Yang\authorrefmark{1,2}, Member, IEEE, and Abdallah Shami\authorrefmark{2}, Fellow, IEEE}
\affil{Faculty of Business and Information Technology, Ontario Tech University, Oshawa, ON L1G 0C5, Canada.}
\affil{Department of Electrical and Computer Engineering, Western University, London, ON N6A 3K7, Canada.}
\corresp{Corresponding authors: Li Yang (email: li.yang@ontariotechu.ca); Abdallah Shami (email: abdallah.shami@uwo.ca).}

\begin{abstract}
With increasingly sophisticated cybersecurity threats and rising demand for network automation, autonomous cybersecurity mechanisms are becoming critical for securing modern networks. The rapid expansion of Internet of Things (IoT) systems amplifies these challenges, as resource-constrained IoT devices demand scalable and efficient security solutions. In this work, an innovative Intrusion Detection System (IDS) utilizing Automated Machine Learning (AutoML) and Multi-Objective Optimization (MOO) is proposed for autonomous and optimized cyber-attack detection in modern networking environments. The proposed IDS framework integrates two primary innovative techniques: Optimized Importance and Percentage-based Automated Feature Selection (OIP-AutoFS) and Optimized Performance, Confidence, and Efficiency-based Combined Algorithm Selection and Hyperparameter Optimization (OPCE-CASH). These components optimize feature selection and model learning processes to strike a balance between intrusion detection effectiveness and computational efficiency. This work presents the first IDS framework that integrates all four AutoML stages and employs multi-objective optimization to jointly optimize detection effectiveness, efficiency, and confidence for deployment in resource-constrained systems. Experimental evaluations over two benchmark cybersecurity datasets demonstrate that the proposed MOO-AutoML IDS outperforms state-of-the-art IDSs, establishing a new benchmark for autonomous, efficient, and optimized security for networks. Designed to support IoT and edge environments with resource constraints, the proposed framework is applicable to a variety of autonomous cybersecurity applications across diverse networked environments.

\end{abstract}

\begin{IEEEkeywords}
Network Automation, AutoML, Multi-Objective Optimization, Cybersecurity, Intrusion Detection System, IoT.
\end{IEEEkeywords}


\maketitle

\section{Introduction}

\IEEEPARstart{T}he fifth generation of wireless networks (5G) has marked a significant step in high-speed data transmission and low latency, which are essential for current and emerging Internet of Things (IoT) applications \cite{6g1}. The Internet of Things (IoT) denotes a network of networked devices that can gather, process, and exchange data, facilitating a system of intelligent and communicative devices \cite{6g2}. IoT technologies have enabled various applications, such as smart homes, smart healthcare, smart agriculture, and intelligent transportation systems \cite{eaai}. Although IoT systems offer numerous benefits, their security and reliability are crucial for their extensive use \cite{ioteffi}.

IoT systems expand the attack surface and introduce new cybersecurity threats due to their connectivity and resource constraints. The consequences of these cyber threats include disruptions of service, financial losses, and risks to human safety \cite{sec1}. Therefore, it is crucial to develop robust cybersecurity mechanisms to protect IoT systems. Ensuring cybersecurity requires the development of Intrusion Detection Systems (IDSs) that detect abnormal behaviors or malicious attacks in network traffic and system activities \cite{mth}. In addition to traditional cybersecurity mechanisms, like encryption and authentication, IDSs provide an additional layer of security by proactively monitoring network traffic and identifying cyber threats in modern networks and IoT systems \cite{tnsm}. 

Artificial Intelligence (AI) and Machine Learning (ML) techniques are effective methods for developing IDSs for protecting networks and IoT systems. ML is a set of data analytics algorithms that enable machines to learn from data and make informed decisions based on that data \cite{thesis}. With network data analytics, AI/ML models can gain insights into network behaviors and detect data patterns that indicate cyber threats. However, traditional ML models require extensive manual configuration and tuning to effectively detect and mitigate threats, making IDS development increasingly challenging as modern networks and IoT systems become more complex and scalable \cite{eaai}.

On the other hand, network automation, also referred to as automated network management, plays a pivotal role in managing the increased complexity and scale of modern networks, while ensuring optimal network performance and security. Zero-Touch Networks (ZTNs) and advanced automation capabilities are envisioned to be vital characteristics of next-generation networks \cite{zsm1}. Therefore, Automated Machine Learning (AutoML) techniques have emerged as a promising technology for enhancing the effectiveness and automation of ML-based IDSs in modern and future networks \cite{ccs} \cite{automl1}. AutoML streamlines the process of developing ML models by automating the data pre-processing, feature engineering, model selection, and hyperparameter tuning of the general data analytics or ML learning process, thereby enabling non-experts to deploy advanced IDSs or other data-driven cybersecurity mechanisms \cite{automl2}. The evolution of AutoML is essential in next-generation networks that demand high levels of automation, as the large amount of data and the dynamic nature of networking environments make continuous manual development and supervision challenging and infeasible \cite{eaai}.

Another key challenge in IDS deployment is achieving a balance between effectiveness and efficiency. The operational requirements of IoT systems, such as energy efficiency, computational constraints, and the need for real-time processing, demand that IDSs not only be accurate but also resource-efficient \cite{ioteffi}. Although AutoML techniques enable the automated development of high-performance cybersecurity mechanisms in IoT and next-generation networks, the development and optimization process of ML models still requires large computational resources \cite{eaai}. Developing cost-effective and resource-efficient cybersecurity solutions is crucial for IoT systems with limited resources, as it facilitates the widespread deployment of security measures and democratizes access to ML-based security models across diverse regions. Multi-Objective Optimization (MOO) complements AutoML by addressing the trade-offs in IDS, such as the balance between detection accuracy and computational overhead \cite{moo1}. This is particularly important in IoT networks, where many devices have limited computational resources and energy constraints. Moreover, MOO provides a flexible framework for optimizing multiple conflicting objectives, enabling the development of IDS that are tailored to the specific constraints and requirements of IoT and modern networks \cite{moo1}.

\begin{figure}
     \centering
     \includegraphics[width=8.5cm]{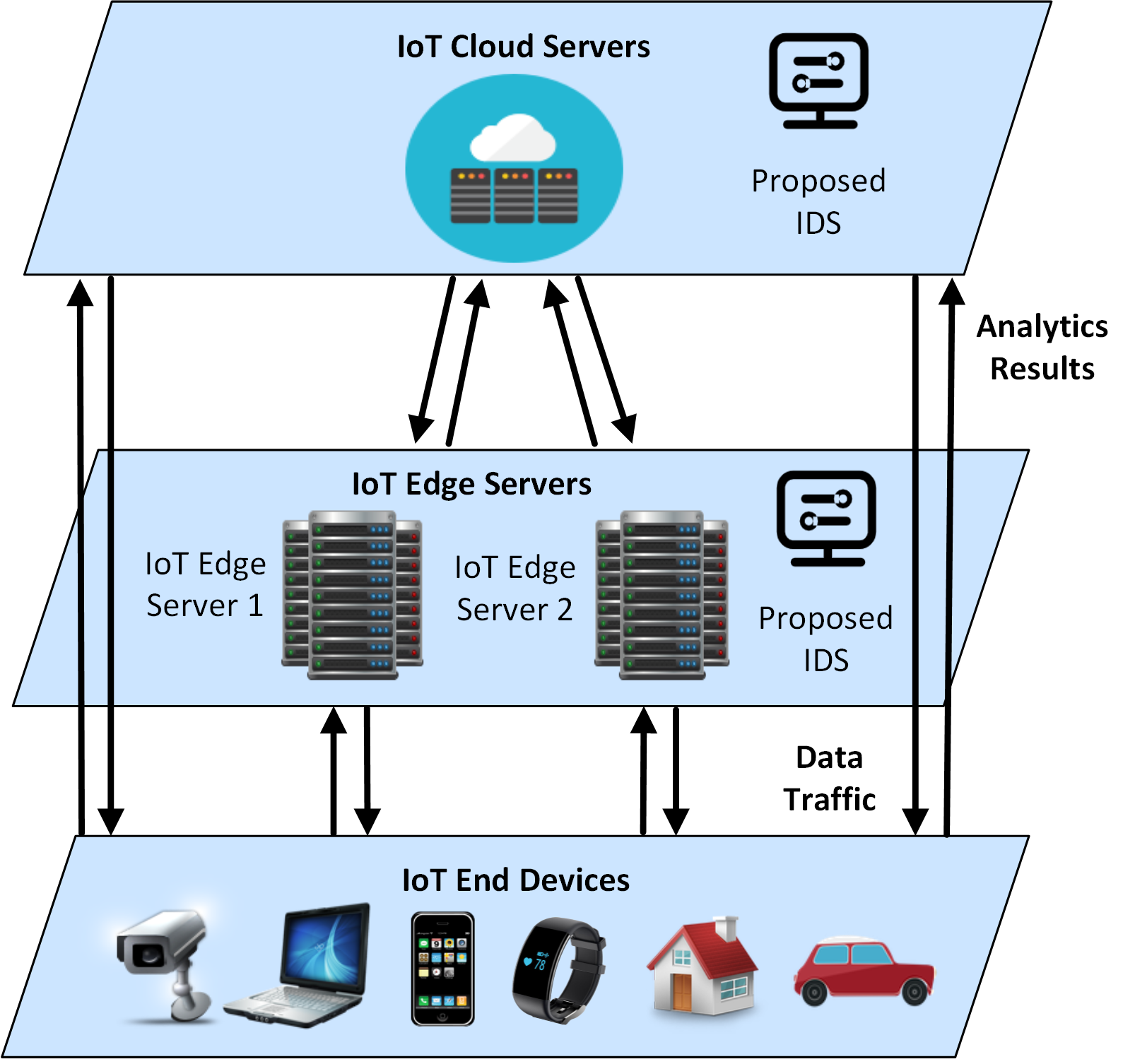}
     \caption{The IoT data analytics architecture and IDS deployment.} \label{iot_moo}
\end{figure}

Therefore, this paper proposes MOO-AutoML IDS, an innovative AutoML and MOO-based IDS to provide effective, efficient, and autonomous cyber-attack detection for modern networks and IoT systems. The proposed framework consists of three stages: Automated Data Pre-Processing (AutoDP), Automated Feature Engineering (AutoFE), and automated model learning and optimization \cite{eaai}. For data quality improvement, the AutoDP process focuses on automated data balancing using the integration of Synthetic Minority Over-sampling Technique (SMOTE) and Adaptive Synthetic Sampling (ADASYN), as well as automated normalization that automatically selects min-max and Z-score methods. In AutoFE, a Multi-Objective Particle Swarm Optimization (MOPSO) based feature selection method, named Optimized Importance and Percentage-based Automated Feature Selection (OIP-AutoFS), is proposed to optimize the accumulated feature importance and the percentage of selected features for generating an optimal feature set for intrusion detection. In automated model learning and optimization, a novel Optimized Performance, Confidence, and Efficiency-based Combined Algorithm Selection and Hyperparameter Optimization (OPCE-CASH) method is proposed for autonomous and efficient intrusion detection. In OPCE-CASH, two tree-based ML models, XGBoost and LightGBM, are trained and optimized using MOPSO, balancing F-score, prediction confidence, and model execution time. The proposed framework is evaluated on two benchmark IoT cybersecurity datasets, CICIDS2017 \cite{cicdata} and IoTID20 \cite{iotid}. As shown in Fig. \ref{iot_moo}, the proposed IDSs can be deployed on IoT cloud servers for large-scale network data analytics and performance-prioritized IDS model training, as well as on edge servers/devices for local data analytics and efficiency-prioritized IDS model predictions \cite{tii}.

This paper makes the following primary contributions:
\begin{enumerate}
\item It proposes MOO-AutoML IDS\footnote{Code is available at: \url{https://github.com/Western-OC2-Lab/Multi-Objective-Optimization-AutoML-based-Intrusion-Detection-System}}, a novel autonomous cybersecurity framework that optimally balances effectiveness and efficiency across all key procedures of ML-based intrusion detection, particularly for resource-constrained environments like IoT systems. \item It proposes OIP-AutoFS, a novel MOO-based automated feature selection component in the MOO-AutoML IDS framework, for obtaining an optimal feature set that optimizes both feature importance and size, thereby striking a balance between model performance and execution time.
\item It proposes OPCE-CASH, a novel MOO-based method for automated IDS model selection and optimization by automatically identifying the best model among optimized XGBoost and LightGBM, balancing effectiveness, reliability, and efficiency.
\item It evaluates the proposed MOO-AutoML IDS on two public benchmark IoT cybersecurity datasets (CICIDS2017 and IoTID20) and compares its performance with state-of-the-art ML and AutoML-based IDSs.
\end{enumerate}

To the best of our knowledge, this is the first research that proposes such an autonomous, optimized, and resource-efficient framework that balances effectiveness, reliability, and efficiency in intrusion detection for modern networks and resource-constrained systems, leveraging both MOO and AutoML techniques.

The manuscript is organized as follows: Section II reviews existing work on AI/ML, MOO, and AutoML for developing IDSs for modern networks. Section III describes the proposed MOO-AutoML IDS framework in detail, including the proposed AutoDP, OIP-AutoFE, and OPCE-CASH methods. Section IV discusses the results of evaluating the proposed framework on benchmark cybersecurity datasets. Section V concludes the key findings of this study.

\section{Related Work}
The rapid advancement of AI/ML techniques has significantly contributed to the evolution of IDSs for modern networks. In this section, a comprehensive literature review is conducted to explore existing AI/ML-based IDSs, from traditional ML-based IDSs to state-of-the-art optimized and automated IDSs for IoT systems and 5G networks. 

\subsection{Traditional ML-based IDSs}
Traditional ML-based IDSs have set the foundation for modern cybersecurity mechanisms, and have been developed in many research works. Sharafaldin \textit{et al.} \cite{cicdata} created an industry-standard cybersecurity dataset called CICIDS2017, which met real-world network criteria, and evaluated seven basic ML algorithms. Fernando \textit{et al.} \cite{ml1} proposed a class-balanced, Dynamically Weighted Balanced (DWB) loss function for cyber intrusion detection, and applied it to traditional ML models and 1-Dimensional Convolutional Neural Networks (1D-CNN). The DWB-based IDS achieved high performance on the CICIDS2017 and ISIC2019 datasets.  

Tree-based ML algorithms are widely used in IDS development because of their interpretability, efficiency, and ability to deal with complex decision boundaries. Bakro \textit{et al.} \cite{ml2} proposed a Random Forest (RF)-based IDS based on the optimal features generated by a hybrid feature selection approach using the Grasshopper Optimization Algorithm (GOA) and the Genetic Algorithm (GA). The RF-based IDS exhibited higher performance than many other ML and Deep Learning (DL) models on benchmark datasets. Mhamdi \textit{et al.} \cite{ml3} proposed a hybrid IDS that combines a deep autoencoder and RF models. The autoencoder model can generate a better representation of the original network traffic dataset and define the normal network traffic behaviors, and the RF model can then identify malicious traffic. This model achieved high anomaly detection rates on Software Defined Networking (SDN) security datasets for next-generation cybersecurity. 
Yang \textit{et al.} \cite{lccde} proposed a novel ensemble model, Leader Class and Confidence Decision Ensemble (LCCDE), to detect cyber-attacks in the Internet of Vehicles (IoV) systems. It integrates three ML algorithms, LightGBM, XGBoost, and CatBoost, based on their performance and confidence values on each type of attack, to construct a reliable IDS that can effectively detect various types of attacks. 

Although many traditional ML-based IDSs have achieved high performance in research projects, it is infeasible or challenging to deploy them into real networks and IoT systems due to resource constraints, especially DL models that require extensive computing resources. Therefore, the performance and efficiency of many traditional IDSs still have room for improvement. Additionally, extensive human supervision and expertise are required in the ML model development process, making it difficult to deploy them in many IoT systems and networking environments.

\subsection{Optimized and Automated IDSs}

\begin{table*}[tbp]
\caption{Comparison of State-of-the-Art ML, Optimized ML, and AutoML-Based IDS Frameworks in the Literature (\checkmark, \raisebox{0.7ex}{$\sqrt{}\mkern-9mu{\smallsetminus}$}, and blank indicate automated implementation, manual implementation (no automation), and non-consideration of the procedure/objective, respectively).}
\label{table:comparison}
\centering
\setlength\extrarowheight{2pt}
\scalebox{0.87}{
\begin{tabular}{|>{\centering\arraybackslash}m{12em}|>{\centering\arraybackslash}m{7em}|>{\centering\arraybackslash}m{7em}|>{\centering\arraybackslash}m{7em}|>{\centering\arraybackslash}m{8em}|>{\centering\arraybackslash}m{6em}|>{\centering\arraybackslash}m{6em}|>{\centering\arraybackslash}m{6em}|}
\hline
\multirow{2}{*}{\textbf{Paper}} & \multicolumn{4}{c|}{\textbf{ML or AutoML Procedures}} & \multicolumn{3}{c|}{\textbf{Optimization Objectives}} \\ \cline{2-8}
& \textbf{Data Pre-Processing} & \textbf{Feature Engineering} & \textbf{Model Selection} & \textbf{Hyperparameter Optimization} & \textbf{Optimized Performance} & \textbf{Optimized Efficiency} & \textbf{Optimized Confidence} \\ \hline 
Sharafaldin \textit{et al.} \cite{cicdata}  & & & \raisebox{0.7ex}{$\sqrt{}\mkern-9mu{\smallsetminus}$} & & & & \\ \hline
Fernando \textit{et al.} \cite{ml1} & \raisebox{0.7ex}{$\sqrt{}\mkern-9mu{\smallsetminus}$} & \raisebox{0.7ex}{$\sqrt{}\mkern-9mu{\smallsetminus}$} & \raisebox{0.7ex}{$\sqrt{}\mkern-9mu{\smallsetminus}$} & & & & \\ \hline
Bakro \textit{et al.} \cite{ml2} & \raisebox{0.7ex}{$\sqrt{}\mkern-9mu{\smallsetminus}$} & \raisebox{0.7ex}{$\sqrt{}\mkern-9mu{\smallsetminus}$} & & & & & \\ \hline
Mhamdi \textit{et al.} \cite{ml3} & & \raisebox{0.7ex}{$\sqrt{}\mkern-9mu{\smallsetminus}$} & & & & & \\ \hline
Yang \textit{et al.} \cite{lccde} & & & \raisebox{0.7ex}{$\sqrt{}\mkern-9mu{\smallsetminus}$} & & & & \\ \hline
Khan \textit{et al.} \cite{oeids} & \raisebox{0.7ex}{$\sqrt{}\mkern-9mu{\smallsetminus}$} & \raisebox{0.7ex}{$\sqrt{}\mkern-9mu{\smallsetminus}$} & \checkmark & \checkmark & \checkmark & & \\ \hline
Singh \textit{et al.} \cite{automlid} & & & \checkmark & \checkmark & \checkmark & & \\ \hline
Brown \textit{et al.} \cite{mal} & & & \checkmark & \checkmark & \checkmark & & \\ \hline
Yang \textit{et al.} \cite{icc} & \raisebox{0.7ex}{$\sqrt{}\mkern-9mu{\smallsetminus}$} & & \raisebox{0.7ex}{$\sqrt{}\mkern-9mu{\smallsetminus}$} & \checkmark & \checkmark & & \\ \hline
Elmasry \textit{et al.} \cite{psolstm} & & \checkmark & & \checkmark & \checkmark & & \\ \hline
Fatani \textit{et al.} \cite{iotop1}  & & \checkmark & & \raisebox{0.7ex}{$\sqrt{}\mkern-9mu{\smallsetminus}$} & \checkmark & & \\ \hline
Sakr \textit{et al.} \cite{iotop2} & & \checkmark & & & \checkmark & & \\ \hline
Qaddos \textit{et al.} \cite{iotop3} & \raisebox{0.7ex}{$\sqrt{}\mkern-9mu{\smallsetminus}$} & \checkmark & & \raisebox{0.7ex}{$\sqrt{}\mkern-9mu{\smallsetminus}$} & \checkmark & & \\ \hline
Subramani \textit{et al.} \cite{svm} & & \checkmark & & & \checkmark & & \\ \hline
Habib \textit{et al.} \cite{moo2} & & \checkmark & & & \checkmark & \checkmark & \\ \hline
Alweshah \textit{et al.} \cite{moo1} & & \checkmark & & & \checkmark & \checkmark & \\ \hline
Our Previous Work \cite{ccs} & \checkmark & \checkmark & \checkmark & \checkmark & \checkmark & & \\ \hline
\textbf{Proposed Framework} & \checkmark & \checkmark & \checkmark & \checkmark & \checkmark & \checkmark & \checkmark \\ \hline
\end{tabular}
}
\end{table*}

As traditional IDSs that use manually-selected ML models with default hyperparameters often face challenges in achieving optimal performance in cybersecurity applications, several autonomous and optimized IDSs have been developed to improve ML model performance using optimization methods.

Despite AutoML being a new technology in cybersecurity applications, there are already several existing papers that develop and enhance IDS using AutoML techniques. Khan \textit{et al.} \cite{oeids} proposed Optimized Ensemble IDS (OE-IDS), an AutoML-based IDS for general networks. The AutoML framework is used to select the most accurate base classifiers, and a soft-voting ensemble method is used to integrate them.  Singh \textit{et al.} \cite{automlid} introduced AutoML-ID, a framework enabling the development of IDS in wireless sensor networks using AutoML. As part of the AutoML-ID method, Bayesian Optimization (BO) is used to automatically select and optimize a model based on the best-performing one among seven given ML models. Brown \textit{et al.} \cite{mal} proposed an AutoML-driven approach for real-time malware detection in cloud environments, which demonstrated improved adaptability to evolving threats. Using AutoML as a solution for automating model selection and optimization in this work outperforms existing CNN-based malware detection models, particularly in noisy environments, demonstrating its adaptability to real-world cybersecurity challenges. Yang \textit{et al.} \cite{ccs} proposed an AutoML-based IDS for 5G/6G networks, which uses a Tabular Variational Auto-Encoder (TVAE) for data pre-processing, tree-based ML models for automated feature selection, and Bayesian Optimization for Hyper-Parameter Optimization (HPO). This method aims to optimize intrusion detection accuracy and outperforms several state-of-the-art methods on the benchmark cybersecurity datasets. 

Existing optimized ML and AutoML-based IDSs can automatically improve ML model performance, but these IDSs only focus on automated model selection and HPO. However, there are other critical components of the AutoML pipeline that should also be considered to further enhance model performance, such as automated data pre-processing and automated feature engineering \cite{eaai}. 
Yang \textit{et al.} \cite{icc} proposed an optimized Convolutional Neural Network (CNN) based IDS for the Internet of Vehicles (IoV) system security. It employs state-of-the-art CNN models and enhances their performance by HPO using the PSO method, as well as model ensemble involving concatenation and confidence averaging methods. This method can achieve better performance than those AutoML methods that only focus on HPO. Elmasry \textit{et al.} \cite{psolstm} proposed a Particle Swarm Optimization (PSO)-based IDS targeting both feature selection and hyperparameter optimization, and applied it to deep learning models, including Deep Neural Networks (DNNs), Long Short-Term Memory (LSTM), and Deep Belief Networks (DBNs). 
 
Fatani \textit{et al.} \cite{iotop1} proposed an efficient intrusion detection framework for IoT environments by coupling convolutional neural networks with a custom Transient Search Optimization Differential Evolution (TSODE) feature selection algorithm. The TSODE method leverages differential evolution to enhance TSO’s convergence, and shows its effectiveness in constrained and evolving IoT settings. They also manually tuned the hyper-parameters of CNNs to improve the detection performance.
Sakr \textit{et al.} \cite{iotop2} proposed a feature selection-driven optimization method to improve IDS efficiency by applying filter, wrapper, and hybrid feature selection techniques, such as GA, Artificial Bee Colony (ABC), and PSO. They compared these feature selection methods using a Support Vector Machine (SVM) as the detection model on an intrusion detection dataset, reinforcing the importance of feature subset optimization for scalable and efficient IDS.
Qaddos \textit{et al.} \cite{iotop3} proposed a hybrid CNN–Gated Recurrent Unit (GRU) intrusion detection framework for IoT environments, integrating a PSO-based automated feature selection method with Feature-Weighted (FW) SMOTE to address class imbalance. The hybrid IDS model demonstrated high adaptability and accuracy on the IoTID20 dataset and outperformed existing IoT-enabled IDS solutions.

However, all the previously mentioned optimized and AutoML-based methods in this section are Single-Objective Optimization (SOO) methods that solely prioritize model accuracy or detection rate optimization. These approaches pose challenges when deploying them in IoT systems or networks with resource constraints or real-time requirements. Additionally, certain network systems deployed with IDSs are sensitive to false alarms or false negatives \cite{svm}. Therefore, other factors/properties of ML models, such as model complexity and inference time, should be considered for IDS development in specific networking environments or cybersecurity applications. 

MOO methods are then developed to enable the optimization and trade-off of multiple factors in cybersecurity frameworks. Subramani \textit{et al.} \cite{svm} proposed a MOPSO-based feature selection method for IDS development in IoT systems. This research optimizes the SVM classifier by improving the detection accuracy while reducing the false positive rate, and evaluates the MOO-based IDS performance on the KDD’99 dataset. However, this method does not consider model complexity. In terms of model complexity or efficiency, Habib \textit{et al.} \cite{moo2} and Alweshah \textit{et al.} \cite{moo1} proposed MOPSO-Lévy and Emperor Penguin Colony (EPC) methods, respectively, to develop multi-objective IDSs for IoT systems. The MOPSO-Lévy and EPC methods are used to achieve a trade-off between accuracy and feature size for optimized feature selection. While these two approaches address both model performance and efficiency in the feature selection process, the most critical process—the model learning process—remains unoptimized in the existing MOO-based IDS literature, resulting in an incomplete MOO process.

\subsection{Literature Comparison and Our Contributions}

To summarize, although there are many existing studies \cite{ccs} \cite{moo1} \cite{cicdata} \cite{ml1} - \cite{moo2} on IDS development for modern networks and IoT systems, they all have certain limitations. 
Several studies \cite{oeids} - \cite{icc} have contributed to the development of IDSs using traditional ML models. They implemented manual data pre-processing, feature engineering, or model selection, but they did not incorporate any important AutoML procedures.
Papers \cite{cicdata} \cite{ml1} - \cite{lccde} have explored AutoML or optimized ML models for IDS development, but their focus has primarily been on either HPO or automated model selection, with limited attention to the integration of AutoML procedures comprehensively. In contrast, papers \cite{psolstm} - \cite{svm} have made strides in automated feature engineering. However, apart from our previous work \cite{ccs}, none of these studies have integrated all key components of AutoML—automated data pre-processing, automated feature engineering, automated model selection, and HPO—into a single IDS framework. Although both our previous work \cite{ccs} and this study have integrated all four key AutoML components, the specific components, algorithms used, and novel contributions differ significantly. Our previous work \cite{ccs} focused on ML model optimization and ensembling for improving model effectiveness, primarily using the F1-score metric, while the proposed work focuses on novel MOO-based automated feature selection and CASH procedures for optimizing model effectiveness, efficiency, and reliability (confidence). Therefore, the current model is more suitable for practical deployment in low-resource environments, such as edge gateways or microcontroller-class IoT devices, which was not the focus of the previous study \cite{ccs}.

Moreover, the objectives of optimizing performance and efficiency have been addressed separately in the existing literature. Papers \cite{moo1} \cite{moo2} have shown significant progress in optimizing performance through multiple objectives. However, none of them considers model reliability and confidence. Additionally, automatically optimizing multi-objectives for IDS development using AutoML techniques has not been achieved until now.

Therefore, the proposed framework integrates both MOO and AutoML techniques to develop a fully autonomous, optimized, and efficient IDS for IoT systems and modern networks. It involves a novel MOO-based automated feature selection method, OIP-AutoFS, which optimizes both feature importance and size. This method strikes a balance between model performance and execution time, addressing a critical need in the FS field. Additionally, we propose OPCE-CASH, an innovative MOO-based method for automated IDS model selection and optimization. Additionally, it proposes automated normalization and data balancing methods for AutoDP that enable data quality improvement. No prior research has presented such a framework that balances effectiveness, reliability, and efficiency in intrusion detection for modern networks and IoT systems using both MOO and AutoML techniques.

\section{Proposed Framework}

\subsection{System Overview}

\begin{figure}[!t]
     \centering
     \includegraphics[width=\columnwidth]{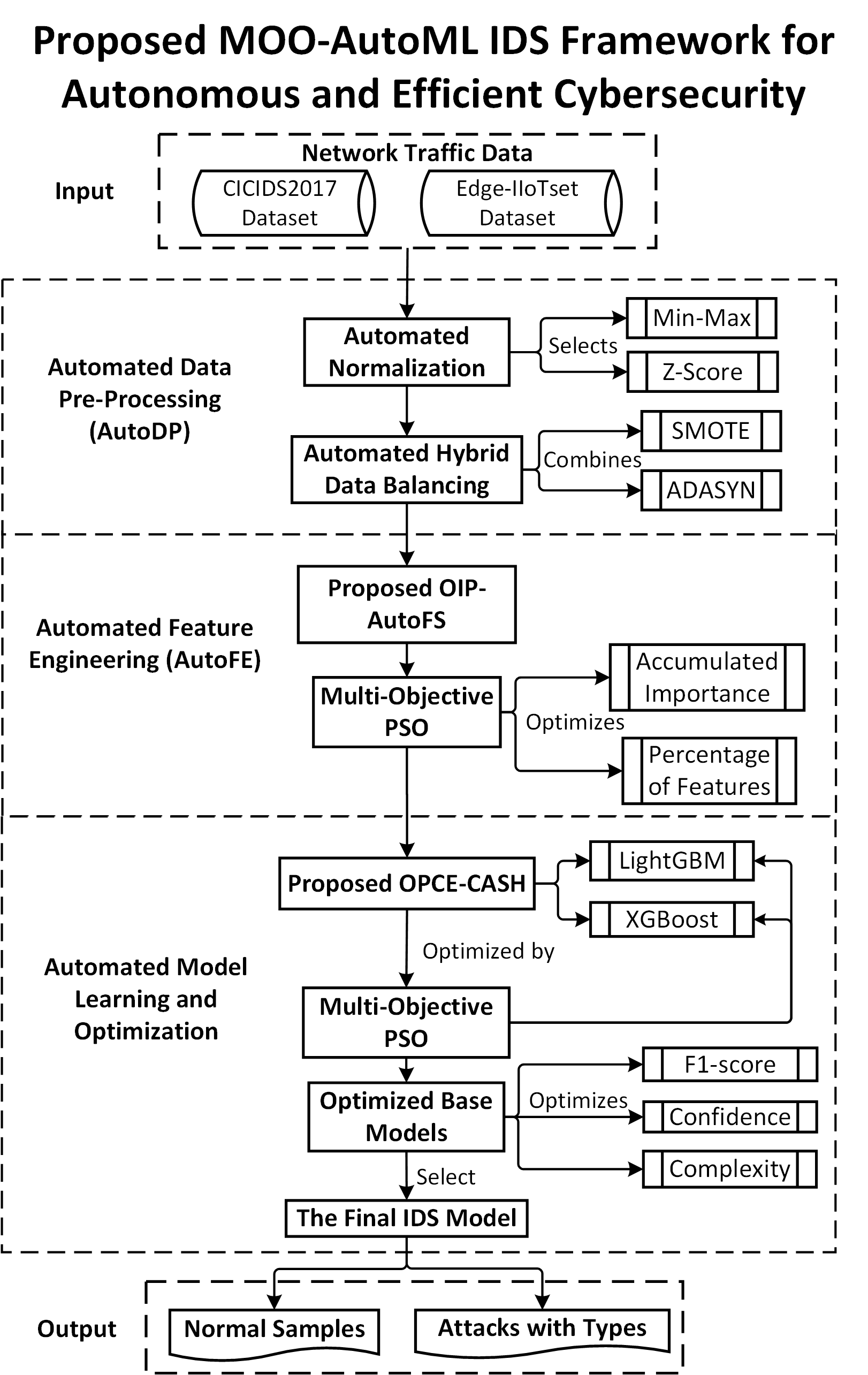}
     \caption{The proposed IDS system overview.} \label{overview}
\end{figure}

The system architecture of the proposed MOO-AutoML IDS for autonomous and efficient cybersecurity is illustrated in Fig. \ref{overview}. The framework is composed of an integrated sequence of automated network data analytics processes designed to ensure optimal and efficient intrusion detection for resource-constrained systems like IoT. The general intrusion detection process begins with collecting network data, and the proposed framework is developed and evaluated using two benchmark cybersecurity datasets: the CICIDS2017 dataset \cite{cicdata} for modern network traffic, and the IoTID20 dataset \cite{iotid} tailored for edge computing scenarios in IoT.

In the initial stage, raw input data is pre-processed using Automated Data Pre-processing (AutoDP) to improve data quality. It involves automated normalization that utilizes two distinct methods: min-max and Z-score, which normalize data variance to improve model fairness. This is complemented by an automated hybrid data balancing approach integrating SMOTE and ADASYN techniques to address class imbalance issues. Subsequently, the framework employs an innovative feature selection algorithm called OIP-AutoFS, which is driven by a Multi-Objective Particle Swarm Optimization (MOPSO) and feature importance. The main purpose of this AutoFS process is to optimize the accumulated importance and percentage of features, thus ensuring that the most significant features are used for model training. 
The automated model learning and optimization using the proposed OPCE-CASH method is the final phase, a critical step where XGBoost and LightGBM are automatically optimized and selected using the MOPSO model. This phase aims to optimize the F1-score (representing model effectiveness), confidence values (representing model reliability), and execution time (indicating model complexity). 

As a final output, the optimal ML model classifies network traffic data into normal samples and various types of attacks, which allows effective and efficient intrusion detection for IoT systems and modern networks. Overall, the proposed MOO-AutoML framework automates all ML procedures to provide an autonomous cybersecurity solution for modern networks. The feature selection and model learning processes are optimized by MOPSO to balance model effectiveness and efficiency, making it well-suited for resource-constrained systems like IoT. The feature selection and model learning procedures have been optimized by MOPSO to achieve a balance between model effectiveness and efficiency, making it well-suited for resource-constrained systems like IoT. The following subsections provide further information on each component of the proposed system. The domain-specific adaptations of the proposed framework’s components to address cybersecurity challenges in modern networks are summarized in Table \ref{components}.

\begin{table*}
\centering
\caption{Domain-Specific Network Challenges, Custom Adaptations, and Performance Impact of Each Component in the Proposed Framework}
\setlength\extrarowheight{2pt}  
\scalebox{0.87}{
\begin{tabular}{|>{\centering\arraybackslash}m{6em}|
                >{\centering\arraybackslash}m{9em}|
                >{\raggedright\arraybackslash}m{18em}|
                >{\raggedright\arraybackslash}m{18em}|
                >{\raggedright\arraybackslash}m{13em}|}
\hline
\textbf{Component} 
& \textbf{Algorithms} 
& \textbf{Network Data/Model Challenges (Cybersecurity/IoT)} 
& \textbf{Adaptation or Customization} 
& \textbf{Impact on IDS Performance} \\
\hline

\multirow{2}{*}{AutoDP} 
& Shapiro–Wilk + Z-score / Min-Max Normalization
& Features in network traffic vary in scale and distribution due to protocol heterogeneity. Manual scaling is error-prone.
& Uses the Shapiro–Wilk test to automatically select the appropriate normalization method between Z-score and min-max normalization.
& Improves model fairness, effectiveness, and generalization across datasets. \\ 
\cline{2-5}

& Hybrid SMOTE + ADASYN 
& Imbalanced class distributions in cybersecurity data, especially rare attacks, impact detection performance.
& Applies SMOTE or ADASYN adaptively, based on class thresholds, to achieve automated data balancing.
& Improves recall of minority attacks and model effectiveness. \\
\hline

AutoFS (OIP-AutoFS) 
& IG + Multi-Objective PSO 
& High-dimensional, redundant features from network data increase cost and reduce robustness.
& Jointly optimizes feature importance and subset size using normalized MOPSO.
& Reduces dimensionality and improves model efficiency and robustness. \\
\hline

OPCE-CASH 
& XGBoost, LightGBM + Multi-Objective CASH (F1, latency, confidence) 
& Resource-constrained systems (\textit{e.g.}, IoT) require ML models that are not only accurate but also fast and trustworthy.
& Simultaneously tunes hyperparameters to optimize F1-score, inference latency, and confidence of ML models.
& Enables real-time, resource-efficient, effective, and reliable detection on edge devices. \\
\hline

AutoML (Overall) 
& Full AutoML Stack (AutoDP, AutoFS, OPCE-CASH) 
& Modern network infrastructures (\textit{e.g.}, zero-touch networks, 5G/5G+) require fully autonomous cybersecurity solutions that minimize human intervention and adapt to dynamic environments.
& Automates the entire ML pipeline: data preprocessing, feature selection, model selection, and hyperparameter tuning, reducing manual engineering.
& Supports scalable, autonomous, and democratic deployment in modern and next-generation networks. \\
\hline

MOO (Overall) 
& MOPSO Framework 
& Resource-constrained network security solutions (\textit{e.g.}, in IoT and edge networks) must strike a balance between multiple objectives, particularly model effectiveness and complexity. Existing IDSs typically optimize one objective only.
& Introduces a unified MOO engine that jointly optimizes model effectiveness (F1-score), efficiency (latency), and reliability (confidence). Applied in both feature selection and model learning stages.
& Enables deployment-aware IDS optimization, ensuring high effectiveness, reliability, and efficiency under resource constraints. \\
\hline

\end{tabular}
}
\label{components}
\end{table*}

\subsection{Automated Data Pre-Processing}
The AutoDP module is designed to enhance the quality of network traffic data, thereby improving intrusion detection performance. It automates both normalization and class balancing by: (1) applying statistical tests to automatically select the most appropriate normalization method for each feature, and (2) dynamically integrating SMOTE and ADASYN to synthesize minority class samples using a threshold-driven hybrid balancing strategy tailored to the class distribution of each dataset.

\subsubsection{Automated Data Normalization}
In ML model learning, the scale of data features has a significant effect on model performance. ML models typically consider features with larger numerical ranges as more important, which can result in biased decisions \cite{mth}. To mitigate this, automated data normalization is employed in the proposed IDS framework. The normalization process involves scaling feature scales to a uniform range to avoid biased ML models. Among existing normalization techniques, Z-score and min-max normalization methods are widely utilized because of their effectiveness and simplicity \cite{norm1}. 

The Z-score normalization transforms every data point, $x_n$, using the mean and standard deviation of its corresponding feature, denoted by \cite{norm1}:
\begin{equation}
x_{n}=\frac{x-\mu}{\sigma},
\end{equation}
where $x$ represents the original value, $\mu$ denotes its mean, and $\sigma$ represents its standard deviation. The Z-score normalization method is particularly effective with datasets that have a Gaussian distribution, as it centers the data around zero and scales it by the standard deviation \cite{eaai}.

Min-max normalization adjusts data points according to their minimum and maximum values \cite{norm1}:
\begin{equation}
x_{n}=\frac{x-min}{max-min},
\end{equation}
where $x$ is the original value of the feature, while $min$ and $max$ are its minimum and maximum values. With this approach, all features are scaled within a 0-1 range, enabling it to be used with datasets with non-Gaussian distributions. However, a primary limitation of this method is its sensitivity to extreme values, resulting in skewed normalized data.

To enable automated data normalization, the suitable normalization technique is automatically selected according to the characteristics/distributions of the dataset. As part of the proposed framework, automated data normalization begins by assessing the distribution of the dataset using Shapiro-Wilk tests. Shapiro-Wilk test is a standard method that evaluates whether a dataset has a normal distribution using p-values \cite{norm2}. A dataset with a p-value higher than 0.05 follows a Gaussian distribution, and it is automatically normalized with Z-scores. In other cases, min-max normalization is used to scale the non-Gaussian data.

\subsubsection{Automated Data Balancing}
Another common issue with network security data is class-imbalance, where normal/benign state data significantly outnumber cyber-attack data. Class-imbalance can lead to biased ML models that detect normal samples accurately but fail to detect actual cyber threats. Therefore, it is important to balance cybersecurity datasets before ML model training to ensure the reliability and robustness of IDSs \cite{mth}. 

Data balancing techniques can be broadly classified as under-sampling and over-sampling methods \cite{eaai}. Under-sampling mitigates class imbalance by reducing the size of the majority classes, such as normal state data in networking environments. By eliminating samples from the majority class randomly, the random undersampling method balances data by creating a more balanced distribution between classes. While under-sampling can reduce the size of the data to improve model efficiency, it also loses valuable information from the majority classes.

In contrast, oversampling increases the number of samples in minority classes to balance the data. Random over-sampling is a basic oversampling method that increases the minority class samples by randomly selecting and duplicating them. Despite its simplicity, random oversampling may result in over-fitting for minority samples since it may become excessively focused on certain minority samples. To address potential over-fitting in the data balancing process, the Synthetic Minority Over-sampling Technique (SMOTE) \cite{smote} and Adaptive Synthetic Sampling (ADASYN) \cite{adasyn} are two advanced methods that can effectively balance cybersecurity datasets by creating high-quality samples for minority classes.

SMOTE \cite{smote} is a distance-based over-sampling method that synthesizes minority class samples by interpolating between existing data samples in the minority classes. The interpolation process is based on the principle of K-Nearest Neighbors (KNN) \cite{smote2}. Based on each data sample $X$ belonging to the minority class, assuming the data sample's $k$ nearest neighbors are represented by $X_1,X_2,\cdots,X_k$, where $X_i$ is a randomly selected neighbor, a new synthetic instance can be defined as follows \cite{smote3}:
\begin{equation}
X_{n}=X+rand(0,1) *\left(X_{i}-X\right), i=1,2, \cdots, k,
\end{equation}
where $rand(0,1)$ represents a randomly generated number within the range of 0 to 1.

ADASYN \cite{adasyn} is a density-based oversampling method that aims to generate samples that are close to the data samples that are difficult to learn from. The ADASYN algorithm begins by identifying the minority class samples that are difficult to classify, typically those that are closest to the majority classes. For each of these difficult samples, ADASYN calculates a difficulty weight based on the distance between the majority classes. Each instance's nearest neighbor in the minority class is evaluated using difficulty weights, which results in the creation of new samples. By interpolating between existing samples and their neighbors, ADASYN ensures not only that the class distribution is balanced, but that the new synthetic samples reflect the complex relationships between classes \cite{adasyn2}. 

\begin{algorithm}[t]
    {\scriptsize
    \caption{Hybrid Automated Data Balancing Method}
    \label{data_balancing}
    \LinesNumbered
    \KwIn{
    \\\quad $\mathcal{D} = \{X_{train}, y_{train}\}$: Training dataset with feature set $X_{train}$ and label set $y_{train}$.
    }
    \KwOut{
    \\\quad $\mathcal{D'} = \{X'_{train}, y'_{train}\}$: Balanced training dataset.
    }

    \tcp{Step 1: Identify class distribution}
    $\text{class\_counts} \leftarrow \text{Counter}(y_{train})$; \tcp*[f]{Count instances per class}
    $\text{avg\_samples} \leftarrow \frac{\sum \text{class\_counts.values()}}{\text{len}(\text{class\_counts})}$; \tcp*[f]{Compute average samples per class}
    $\text{threshold} \leftarrow \frac{\text{avg\_samples}}{2}$; \tcp*[f]{Define threshold for minority classes}
    
    \tcp{Step 2: Identify minority classes}
    $\text{minority\_classes} \leftarrow \{ \text{ if } \text{class\_counts} < \text{threshold}\}$; 

    \tcp{Step 3: Apply hybrid over-sampling approach}
    \uIf{$\text{minority\_classes} \neq \emptyset$}{
        \tcp{First, apply SMOTE to generate 50\% of required synthetic samples}
        $\text{smote} \leftarrow \text{SMOTE}(\text{sampling\_strategy}=\text{minority\_classes})$; 
        $X_{train}, y_{train} \leftarrow \text{smote.fit\_resample}(X_{train}, y_{train})$; 
        
        \tcp{Then, apply ADASYN to generate the remaining 50\% of synthetic samples}
        $\text{adasyn} \leftarrow \text{ADASYN}( \text{sampling\_strategy}=\text{minority\_classes})$; 
        $X'_{train}, y'_{train} \leftarrow \text{adasyn.fit\_resample}(X_{train}, y_{train})$; 
    }

    \tcp{Step 4: Output the final balanced training dataset}
    \Return $\mathcal{D'} = \{X'_{train}, y'_{train}\}$; 

}
\end{algorithm}

The specific procedures of the proposed hybrid automated data balancing method are shown in Algorithm \ref{data_balancing}. The first step of this method is to automatically identify whether the distribution of data is balanced. This is achieved by computing the average number of samples per class across the dataset, and then identifying classes that have fewer samples than half of this average as minority classes. The rationale behind selecting half the average as the threshold is to ensure a fair representation of classes without overly creating minority class samples. Once the minority classes are identified, the method employs a two-stage approach to generate samples. It uses SMOTE for creating 50\% of the required new samples for minority classes and ADASYN for the remaining 50\%. The choice of a 50\%/50\% split is strategic: it leverages the strengths of both techniques while mitigating their weaknesses. Using SMOTE, synthetic samples can be produced by interpolating between minority samples, but nuanced and complex class boundaries may not be adequately captured. ADASYN complements this by focusing on minority samples that are difficult to learn, resulting in a variety of examples and ensuring the synthetic data represents actual data complexity. Using SMOTE and ADASYN in combination ensures that cybersecurity datasets are more robust and comprehensive when addressing class imbalances.

The proposed hybrid data balancing method has the following advantages: 
\begin{enumerate}
\item Over-sampling methods (\textit{i.e.}, SMOTE and ADASYN) are particularly effective in intrusion detection. They balance data by synthesizing minority attack samples, thereby preserving crucial information in normal samples. 
\item Unlike traditional over-sampling methods, such as random over-sampling that simply duplicates existing samples and may cause over-fitting, SMOTE and ADASYN can generate high-quality minority samples based on distance measures and learning difficulties to construct a more robust ML model and avoid over-fitting. 
\item SMOTE is effective at balancing data with simple distributions, while ADASYN is effective at handling data with difficult distributions. Therefore, the proposed hybrid method can be applied to complex network datasets of varying complexity.
\item Integrating both models allows for flexible tuning of generated samples. The ratio of synthetic samples generated by each method can be tuned to find the optimal balance for specific problems and datasets.
\end{enumerate}

\subsection{Proposed OIP-AutoFS Method for Automated Feature Engineering} \label{fs}

\subsubsection{Introduction to Feature Selection and MOO}
Through the proposed AutoDP process, the original dataset has been balanced and scaled to improve data quality. In the next step, Automated Feature Engineering (AutoFE) is implemented to automatically generate an optimal feature set and further enhance data quality. 

Feature Engineering (FE) is a crucial step in the ML pipeline, which aims to generate an improved feature set that can better reflect the data patterns used for decision-making, as original features in real-world datasets are often not the optimal features for certain tasks \cite{eaai}. The FE process can significantly influence ML models' performance and determine their upper limits \cite{automl2}. 

This research focuses on Feature Selection (FS), a critical component of FE, which removes redundant and irrelevant features from the feature set. This step is especially important for real-world network datasets, since they usually include a large number of attributes/features collected from various devices, tools, and sensors.  For example, a study collected 248 unique network features for network flow classification \cite{fs1}. Excessive features may cause overfitting and extra execution time due to noisy features, and an appropriate FS process can reduce ML model complexity and improve its generalization ability. However, manual FS often requires significant time and effort, as well as domain expertise. Automated FS (AutoFS) is the process of automating the time- and labor-intensive steps in the traditional FS process, thereby enhancing the efficiency and effectiveness of ML models.

On the other hand, researchers have been exploring the application of MOO to FS tasks \cite{moo1} \cite{svm} \cite{moo2}. Optimization is a fundamental process in ML that aims to identify the optimal solution from all feasible solutions. Common optimization procedures in AutoML applications involve selecting the optimal feature set using optimization techniques. MOO, an extension of traditional Single-Objective Optimization (SOO), aims to optimize multiple objectives simultaneously. MOO problems are often more complex than SOO problems, as the multiple objectives can be conflicting.

MOO is implemented in the proposed framework because balancing accuracy with efficiency in FS is critical. When many features are eliminated during feature selection, underfitting can occur as important information is lost. In contrast, keeping redundant or noisy features can negatively impact model performance. To maximize the overall effectiveness and efficiency of the learning model when fed with the optimal set of features, it is necessary to balance prediction performance and computational efficiency. Due to these constraints, MOO presents a promising means of striking a balance between model performance and efficiency, particularly in IoT cybersecurity applications.

\subsubsection{Multi-Objective Particle Swarm Optimization (MOPSO)}
Particle Swarm Optimization (PSO) is a stochastic optimization method based on the collective behavior observed in flocks of birds or schools of fish. Multi-Objective PSO (MOPSO) is an improved version of PSO that facilitates the optimization of multiple objectives rather than a singular objective \cite{svm}. The MOPSO algorithm belongs to the evolutionary algorithm category and is selected for its simplicity, efficiency, and effectiveness in exploring and exploiting the search space in MOO.

The MOPSO process begins with the initialization of a swarm of particles, each representing a potential solution to the optimization problem. In the proposed AutoFS approach, these particles refer to a distinctive combination of characteristics. The initialization phase is identified by randomness, ensuring a diverse range of starting points for the optimization phase.

Following the initialization of particles, the fitness evaluation of each particle is conducted based on a well-defined objective function. In the proposed framework, the objective function aims to achieve an optimal balance between effectiveness and efficiency. This multi-objective approach ensures that the resulting model is both effective in detection capabilities and efficient in computational resource utilization.

The core mechanism of MOPSO is the continuous update of every particle's position and velocity. The update of the position is a critical operation, as it signifies a transition towards a potentially optimal solution. Meanwhile, the velocity update dictates this transition's trajectory and magnitude. A particle's velocity is recalculated by integrating several factors: the particle's existing velocity, its deviation from the swarm's global best position, which is expressed as:
\begin{equation}
v_{i}^{(t+1)} = w \cdot v_{i}^{(t)} + c_1 \cdot r_1 \cdot (p_{\text{best},i} - x_{i}^{(t)}) + c_2 \cdot r_2 \cdot (g_{\text{best}} - x_{i}^{(t)}) 
\end{equation}
where \( v_{i} \) denotes the particle's velocity, $x_i$ is the particle's position, and \( w \) represents the inertia weight influencing the effect of the preceding velocity. The acceleration coefficients \( c_1 \) and \( c_2 \) modulate the influence of the best positions on the process, while the random variables \( r_1 \) and \( r_2 \) introduce stochastic elements.

Each particle's position is modified after the velocity update. This modification involves the addition of the newly computed velocity to the current position, denoted by:
\begin{equation}
 x_{i}^{(t+1)} = x_{i}^{(t)} + v_{i}^{(t+1)} 
\end{equation}

An updated position indicates a new solution for AutoFS. MOPSO balances the exploration of the search space with the exploitation of the acquired knowledge through the interaction between velocity and position updates. It ensures that MOPSO will pursue the optimal set of features for intrusion detection while maintaining a balance between detection accuracy and computational complexity, while navigating the solution space with efficiency.
 
The computational complexity of MOPSO is $O(nfo)$, where $n$ is the number of particles, $f$ is the number of features, and $o$ is the number of objectives. In the proposed AutoFS method, $o = 2$, while in the proposed CASH method for model learning, $o = 3$. MOPSO is selected as the optimization method in the proposed framework for optimizing both the AutoFS and model learning procedures due to the following primary reasons \cite{svm} \cite{hpome}:
\begin{enumerate}
\item MOPSO works efficiently with high-dimensional hyperparameters or feature spaces, as for instance in the proposed framework, which uses large and complex ML models (tree-based models) for cybersecurity datasets.
\item MOPSO can handle all types of variables, including continuous, discrete, and categorical features. However, many other optimization methods, like gradient-based methods and Bayesian Optimization with Gaussian Process (BO-GP), are only effective with continuous variables \cite{hpome}.
\item MOPSO has a low computational complexity of $O(nfo)$, which is lower than many other optimization methods, such as grid search and random search. 
\item MOPSO can be easily parallelized to improve learning efficiency, as different evaluations can be implemented individually. In contrast, many other optimization methods, such as Genetic Algorithms (GA), are challenging to parallelize due to their sequential nature.
\end{enumerate}

MOPSO is used in both the feature selection and the ML model learning process of the proposed IDS framework, which will be discussed in Sections III-C.\ref{autofs} and III-D.\ref{cash}, respectively.

\subsubsection{Proposed OIP-AutoFS Method} \label{autofs}
To strike a balance between model effectiveness and efficiency, two metrics are optimized by MOPSO in the proposed OIP-AutoFS method: the accumulated feature importance score and the percentage of selected features. The importance of features in ML applications indicates how influential they are in predicting a target variable. The feature importance-based FS methods assign a score to each feature and then select the important features with higher scores. Calculating feature importance scores in IDS development is crucial for identifying key indicators of various cyber-attacks for further analysis and the design of countermeasures \cite{treeme}. 

The proposed OIP-AutoFS module introduces a novel dual-objective formulation for feature selection, which simultaneously optimizes the normalized cumulative Information Gain (IG) of the selected features and minimizes the proportion of features retained. Unlike existing PSO-based feature selection methods \cite{psolstm} \cite{iotop3} that treat objectives independently, the OIP-AutoFS framework uses a normalized trade-off algorithm to prevent objective domination between objectives and achieve a balanced solution during multi-objective optimization. This is particularly important for high-dimensional network traffic data, where feature redundancy can affect IDS generalization on cyber-attacks.

In the proposed AutoFS method, IG \cite{ig} is used to measure the importance of each feature. IG is a concept from information theory that measures the increase or reduction in entropy, quantifying the amount of information a feature provides to the target variable. The IG value of a feature $X$ relative to a target variable $T$, can be determined using the following equation \cite{ig}: 
\begin{equation}
IG(T | X)=H(T)-H(T | X)\label{IGeq},
\end{equation}
where $H(T)$ refers to the entropy of a variable $T$, while $H(T|X)$ refers to the conditional entropy of $T$ relative to $X$. $IG(T|X)$ indicates how important feature $X$ is in relation to target $T$. For example, the feature $X$ has a greater influence on the target variable $T$ if $IG(T|X)$ is greater than $IG(T|Y)$.

The IG method is used for feature importance computation and feature selection as it can directly measure the amount of information a feature contributes to making a decision, providing a clearer and more straightforward indicator than other indirect or composite metrics. Additionally, the IG method has a low computational complexity of $O(n)$ \cite{ig}.

After obtaining the importance scores of all features using the IG method, they are converted into relative values that have a sum of 1.0 or 100\%. This normalization is achieved by dividing each feature's IG score by the sum of all scores. This normalization method ensures that each importance score represents the feature's relative contribution to the total information gained. The first optimization objective, the accumulated feature importance score, is then computed by adding the relative importance scores of all the selected features. 

The percentage of selected features is the second metric considered in the AutoFS process. Balancing the accumulated importance and the percentage of selected features is important in cybersecurity applications with fast decision-making requirements, especially resource-limited IoT systems. This is because keeping a large percentage of features can cause additional execution time and resource consumption, while a small percentage of features may result in a loss of critical information and under-fitting. Reducing the feature set while maintaining a high accumulated importance score can help improve model efficiency. 

Both the accumulated feature importance score and the percentage of selected features are in the range of 0 to 1. Scaling both objectives within the same weight can prevent one metric from dominating the optimization process due to scale differences, thereby avoiding biased results. This is crucial in the MOO process, where the primary objective is to balance multiple objectives. The proposed OIP-AutoFS method can use the MOPSO algorithm to optimize these two variables to select the optimal feature set that can be used to strike a balance between ML model performance and efficiency.

The specific steps of the proposed OIP-AutoFS method are summarized in Algorithm \ref{oip_autofs}. 
In this method, each MOPSO particle encodes a binary vector where each bit represents the inclusion or exclusion of a specific network traffic feature. The particle's position indicates a candidate subset of features, and its velocity determines the probability of switching the selection status for each feature during the next iteration.
In intrusion detection, this enables particles to collaboratively explore various feature combinations to optimize detection performance while reducing computational burden. MOPSO evaluates each particle in a two-objective fitness space: (1) the normalized cumulative importance of selected features (effectiveness), and (2) the normalized percentage of selected features (efficiency), both scaled into the range of 0 to 1. Equal weight is given to both objectives to reflect a balanced trade-off. These weights can be modified for device-specific constraints, but we adopt equal weights to ensure generality across IDS environments.

\begin{algorithm}[t]
{\scriptsize
\caption{Optimized Importance and Percentage-Based Automated Feature Selection (OIP-AutoFS)}
\label{oip_autofs}
\LinesNumbered
\KwIn{
\\\quad $\mathcal{D} = \{X, Y\}$: Dataset with feature set $X$ and target variable $Y$. \\
\quad Each particle encodes a binary feature selection mask over $X$
}
\KwOut{
\\\quad $\mathcal{F}_{opt}$: Optimized feature set.
}

\tcp{Step 1: Initialize the swarm for MOPSO}
$\mathcal{S} \leftarrow \text{InitializeSwarm}()$; \tcp*[f]{Generate initial swarm of particles}

\tcp{Step 2: Compute feature importance scores using Information Gain (IG)}
\ForEach{feature $x_i$ in $X$}{
    $IG_i \leftarrow H(Y) - H(Y|x_i)$; \tcp*[f]{Compute IG for feature $x_i$}
}
$IG_{norm} \leftarrow \frac{IG}{\sum IG}$; \tcp*[f]{Normalize IG scores to sum to 1}

\tcp{Step 3: Multi-Objective Optimization with MOPSO}
\While{termination criteria not met}{
    \ForEach{particle $p$ in $\mathcal{S}$}{

        \tcp{Each particle encodes a binary vector: 1 = selected, 0 = excluded}
        $\mathcal{F}_p \leftarrow \text{SelectFeatures}(p, X)$;

        \tcp{Calculate normalized objective scores}
        $Score_{importance} \leftarrow \sum IG_{norm}[\mathcal{F}_p]$; \tcp*[f]{Effectiveness (normalized)}
        $Score_{percentage} \leftarrow \frac{|\mathcal{F}_p|}{|X|}$; \tcp*[f]{Efficiency (normalized)}

        \tcp{Equal weights are used; adaptable for device-specific constraints}
        $\text{Fitness}_p \leftarrow [Score_{importance}, Score_{percentage}]$;

        \tcp{Update particle position and velocity based on MOPSO logic}
        $\text{UpdateParticle}(p, \text{Fitness}_p)$;
    }

    $\text{UpdateSwarm}(\mathcal{S})$; \tcp*[f]{Update global archive and apply non-dominated sorting}
}

\tcp{Step 4: Extract the optimal Pareto solution}
$\mathcal{F}_{opt} \leftarrow \text{ExtractOptimalFeatures}(\mathcal{S})$; 

\Return{$\mathcal{F}_{opt}$};
}
\end{algorithm}

\subsection{Automated Model Learning and Optimization} \label{learning}
\subsubsection{Base Model Learning}

After implementing the AutoDP and AutoFS procedures, the improved network traffic datasets are trained by ML models to perform intrusion detection. In network intrusion detection problems, the objectives are to distinguish cyber-attacks from normal events and to identify various types of specific attacks. Therefore, intrusion detection problems are multi-class classification problems that can be solved by ML classifiers. In the proposed framework, two tree-based ML models, XGBoost and LightGBM, are used to develop base classifiers to detect multiple types of attacks in IoT systems. 

Tree-based ML algorithms have emerged as effective methods in ML and network data analytics applications \cite{treeme}. A Decision Tree (DT) \cite{dt} is a foundational supervised ML algorithm that divides a dataset into subsets based on its input features, thus creating a tree-structure model of decisions. Their ability to automatically select relevant features, optimize splits using criteria like Gini impurity or entropy, and reduce model complexity using pruning makes them effective and computationally efficient \cite{hpome}. However, DTs are vulnerable to high variance, which can lead to over-fitting issues. To address this, various ensemble learning techniques, such as XGBoost and LightGBM, have been developed to enhance model performance and reduce overfitting.

XGBoost, an advanced tree-based algorithm, is known for its speed and performance enhancements in Gradient-Boosted Decision Trees (GBDTs) \cite{xgboost}. XGBoost employs a second-order Taylor approximation for the summation of squared errors and incorporates a regularization component for minimizing the loss function. Given a differentiable loss function \( \ell(y_i, \hat{y}_i) \), the second-order expansion can be denoted by \cite{xgboost}:
\begin{equation}
    \ell(y_i, \hat{y}_i^{(t-1)} + f_t(\mathbf{x}_i)) \approx \ell(y_i, \hat{y}_i^{(t-1)}) + g_i f_t(\mathbf{x}_i) + \frac{1}{2} h_i f_t^2(\mathbf{x}_i),
\end{equation}
where \( g_i = \frac{\partial \ell(y_i, \hat{y}_i)}{\partial \hat{y}_i} \) is the first-order gradient, and \( h_i = \frac{\partial^2 \ell(y_i, \hat{y}_i)}{\partial \hat{y}_i^2} \) is the second-order gradient. \( f_t(\mathbf{x}_i) \) represents the prediction of the new tree at iteration \( t \). 

Additionally, to control overfitting, XGBoost employs an L2 regularization term that penalizes complex base trees, which is represented by:
\begin{equation}
    \Omega(f_t) = \gamma T + \frac{1}{2} \lambda \sum_{j=1}^{T} w_j^2,
\end{equation}
where \( T \)  is the number of leaf nodes in the base tree, \( w_j \) represents the weight for leaf node \( j \), \( \gamma \) is a parameter to control tree growth, and \( \lambda \) is the L2 regularization parameter.

Using the Second-Order Taylor Approximation and L2 regularization term, the XGBoost model updates trees by optimizing the following loss function at each iteration:
\begin{equation}
    \mathcal{L}^{(t)} = \sum_{i=1}^{n} \left[ g_i f_t(\mathbf{x}_i) + \frac{1}{2} h_i f_t^2(\mathbf{x}_i) \right] + \Omega(f_t).
\end{equation}

XGBoost has an efficient computational complexity of $O(K d|\mathbf{x}| \log n)$, where $K$ represents the number of trees, $d$ indicates the maximum depth of these trees, $|\mathbf{x}|$ indicates the number of non-missing samples, and $n$ is the dataset's overall size \cite{xgboost}. Furthermore, XGBoost supports parallel execution, which significantly reduces model training time and enhances efficiency.

LightGBM is a rapid and robust machine learning model that leverages multiple decision trees for ensemble learning \cite{lightgbm}. It is recognized by its ability to handle large datasets that are high-dimensional and of a large scale. Two key strategies are incorporated in LightGBM: Gradient-based One-Side Sampling (GOSS) and Exclusive Feature Bundling (EFB) \cite{lightgbm}. The GOSS technique is an innovative down-sampling method that prioritizes data samples with larger gradients and discards samples with smaller gradients at random, which can be denoted by:
\begin{equation} \mathcal{L}_{\text{GOSS}} = \sum_{i \in \mathcal{S}_1} \ell(y_i, \hat{y}_i) + \omega \sum_{i \in \mathcal{S}_2} \ell(y_i, \hat{y}_i) \end{equation}
where $\ell(y_i, \hat{y}_i)$ is the loss function, \( \mathcal{S}_1 \) involves the selected high-gradient samples, \( \mathcal{S}_2 \) includes the randomly sampled low-gradient points, and \( \omega \) is a reweighting factor to maintain gradient distribution. Using GOSS speeds up the training process as well as reduces the amount of required memory.

The EFB technique is a feature engineering methodology that consolidates multiple features into a single bundle:
\begin{equation}
    \min \sum_{p=1}^{P} C_p, \quad \text{s.t.} \quad \forall X_i, \quad \sum_{j \in B_p} I(X_{ij} \neq 0) \leq 1
\end{equation}
where \( C_p \) indicates the total feature conflicts in bundle \( p \), \( B_p \) is a feature bundle, and \( I(X_{ij} \neq 0) \) ensures features in a bundle do not co-occur in the same data sample. By utilizing EFB, the overall number of features can be reduced, thereby enhancing the effectiveness of the model training process. 

Combining GOSS and EFB significantly reduces file sizes without compromising important information. Therefore, LightGBM's time and space complexity is effectively reduced to $O(N^\prime F^\prime)$, where $N^\prime$ is the number of samples post-GOSS application, and $F^\prime$ is the number of EFB features \cite{iotm}.

There are several key advantages to selecting XGBoost and LightGBM as classifiers in the proposed framework \cite{thesis} \cite{xgboost} \cite{lightgbm}:
\begin{enumerate}
\item XGBoost and LightGBM are robust ensemble models that have been proven to perform well in a variety of applications, including large-scale network traffic data analytics problems. This is primarily due to their ensemble architecture, which is capable of handling complex data samples during the training process; their regularization feature, which helps to prevent overfitting; and their built-in mechanism for handling missing data.
\item As a result of their training processes, these two tree-based algorithms can generate feature importance scores, which are beneficial for feature selection.
\item Both XGBoost and LightGBM have a low computational complexity for fast execution. XGBoost and LightGBM support parallelization and graphics processing unit (GPU) execution, accelerating the learning process.
\item Both tree-based models inherently incorporate randomness in their construction process. As a result of this randomness, an ensemble model with a high degree of diversity and generalizability can be developed.
\item As the proposed framework targets resource-constrained network environments such as IoT systems, tree-based models (\textit{e.g.}, XGBoost, LightGBM) are selected due to their superior computational efficiency and minimal memory overhead, which render them more suitable for deployment on microcontroller-class devices typical of IoT edge nodes, where deep learning models are often impractical due to excessive latency and resource demands.
\end{enumerate}

\subsubsection{Proposed OPCE-CASH for Automated Model Optimization} \label{cash}

Before ML model training, hyperparameters, which determine the architectures of ML algorithms, must be set \cite{automl2}. As part of the proposed framework, the performance of ML models, including XGBoost and LightGBM, is directly influenced by their hyperparameters.  These tree-based algorithms are subject to several important hyperparameters. First, the number of base DTs to be combined, 'n\_estimators', is a crucial hyperparameter in XGBoost and LightGBM models as it determines the size of the ensemble and, thus, its ability and efficiency to learn from the data. The 'learning\_rate' hyperparameter is also an important hyperparameter that controls the rate of model learning. A slower learning rate requires more trees to model all the relationships, but can potentially result in a more robust model at the cost of increased computational time. Another critical hyperparameter is 'max\_depth', which controls the maximum depth of base trees. This parameter is essential, as it directly influences the complexity of the base DTs in the ensemble models. Deeper trees with more subtrees can make more accurate decisions, but they also carry the risk of overfitting \cite{treeme}. 

As these hyperparameters have a significant impact on ML model performance, it is crucial to tune them for ML performance improvement and ultimately select the optimal model for intrusion detection. However, manual ML model tuning and selection face several key challenges, including time-consuming model development, human bias and errors, and the need for ML expertise, making it difficult to obtain the optimal model \cite{ztnme}. Therefore, the Combined Algorithm Selection and Hyper-parameter optimization (CASH) process is proposed to enable autonomous ML selection and hyperparameter tuning/optimization by using optimization techniques \cite{mirna1}. 

The CASH problem aims to simultaneously determine the most appropriate machine learning algorithm \( \mathcal{A}^{\star} \) and its optimal hyperparameter configuration \( \lambda^{\star} \) that collectively yield the lowest possible loss \( \mathcal{L} \). This process is denoted by:
\begin{equation}
    \mathcal{A}^{\star}, {\lambda}^{\star} \in \underset{\mathcal{A}^{(j)} \in \mathcal{A}, \lambda^{(j)} \in \boldsymbol{\lambda}}{\operatorname{argmin}} \frac{1}{K} \sum_{i=1}^{K} \mathcal{L}\left(\mathcal{A}^{(j)},  \lambda^{(j)}, D_{\text{train}}^{(i)}, D_{\text{valid}}^{(i)}\right)
\end{equation}
where \( \mathcal{A}^{(j)} \in \mathcal{A} \) represents a candidate ML algorithm from a predefined set of possible models (\textit{e.g.}, XGBoost and LightGBM), \( \lambda^{(j)} \in \boldsymbol{\lambda} \) denotes the hyperparameters corresponding to the selected ML algorithm. \( D_{\text{train}}^{(i)} \) and \( D_{\text{valid}}^{(i)} \) represent the training and validation datasets for the \( i \)-th fold in \( K \)-fold cross-validation, and \( K \) is the total number of folds used for model evaluation.

The goal of solving the CASH problem is to efficiently explore the joint space of ML algorithms and hyperparameters to identify the combination that minimizes the validation loss across the cross-validation folds. This approach is fundamental to ML, as it automates the selection of both the learning model and its hyperparameters, thus improving model generalization while democratizing ML development by reducing the need for manual tuning and human expertise.

Similar to the proposed OIP-AutoFS method described in Section III-C, MOPSO is also used in the proposed Optimized Performance, Confidence, and Efficiency-based Combined Algorithm Selection and Hyperparameter Optimization (OPCE-CASH) technique to optimize and select from two base ML algorithms, XGBoost and LightGBM. As described in Section III-C, MOPSO is chosen as the optimization method in the proposed framework due to its ability to handle high-dimensional spaces and diverse variable types with reduced computational complexity and enhanced parallelization capabilities, making it ideal for complex ML models in cybersecurity applications \cite{svm} \cite{hpome}. Additionally, MOPSO allows simultaneous optimization of multiple objectives, including F1-score, confidence/probability, and model complexity (execution/learning time). For IoT security mechanisms, a balance between model performance and complexity is essential due to resource constraints.

\begin{algorithm}[t]
\scriptsize
\caption{Optimized Performance, Confidence, and Efficiency-based Combined Algorithm Selection and Hyperparameter Optimization (OPCE-CASH)}
\label{opce_cash}
\LinesNumbered
\KwIn{
    $\mathcal{D} = \{X, Y\}$: Dataset with feature set $X$ and target labels $Y$. \\
    $\mathcal{A} = \{\text{XGBoost}, \text{LightGBM}\}$: Candidate ML algorithms. \\
    $\lambda$: Set of hyperparameters for each algorithm in $\mathcal{A}$. \\
    $K$: Number of cross-validation folds.
}
\KwOut{
    $\mathcal{A}^{\star}, \lambda^{\star}$: Optimized ML algorithm and hyperparameter configuration.
}

\tcp{Step 1: Initialize the swarm for MOPSO}
$\mathcal{S} \leftarrow \text{InitializeSwarm}()$ \tcp*[f]{Each particle encodes model + hyperparameters}

\tcp{Step 2: Evaluate fitness for each particle}
\ForEach{particle $p$ in $\mathcal{S}$}{

    Decode $\mathcal{A}^{(p)}$ and $\lambda^{(p)}$ from $p$'s position vector; \tcp*[f]{Model + configuration}

    \tcp{Perform K-fold cross-validation}
    \For{$i = 1$ \textbf{to} $K$}{
        Split $\mathcal{D}$ into training set $D_{\text{train}}^{(i)}$ and validation set $D_{\text{valid}}^{(i)}$; \\
        Train $\mathcal{A}^{(p)}$ with $\lambda^{(p)}$ on $D_{\text{train}}^{(i)}$; \\
        Evaluate on $D_{\text{valid}}^{(i)}$ to obtain: \\
        \quad $F1^{(i)} \leftarrow$ F1-score (effectiveness); \\
        \quad $C^{(i)} \leftarrow$ Average prediction confidence (reliability); \\
        \quad $T^{(i)} \leftarrow$ Inference latency (efficiency);
    }

    \tcp{Step 3: Normalize and compute objective vector}
    $F1_{\text{avg}} \leftarrow \frac{1}{K} \sum_{i=1}^{K} F1^{(i)}$; \\
    $C_{\text{avg}} \leftarrow \frac{1}{K} \sum_{i=1}^{K} C^{(i)}$; \\
    $T_{\text{avg}} \leftarrow \frac{1}{K} \sum_{i=1}^{K} T^{(i)}$; \\
    $T_{\text{norm}} \leftarrow \text{Sigmoid}(T_{\text{avg}})$; \tcp*[f]{Normalize latency to [0,1]}

    \tcp{Use equal weights for 3 objectives; reweighting is supported}
    $\text{Fitness}_p \leftarrow [F1_{\text{avg}}, C_{\text{avg}}, T_{\text{norm}}]$;

    \tcp{Step 4: Update particle using MOPSO logic}
    $\text{UpdateParticle}(p, \text{Fitness}_p)$;
}

\tcp{Step 5: Evolve swarm until convergence}
\While{termination criteria not met}{
    $\text{UpdateSwarm}(\mathcal{S})$; \tcp*[f]{Non-dominated sorting + velocity/position update}
}

\tcp{Step 6: Extract optimal configuration}
$\mathcal{A}^{\star}, \lambda^{\star} \leftarrow \text{ExtractOptimalSolution}(\mathcal{S})$; 

\Return{$\mathcal{A}^{\star}, \lambda^{\star}$};

\end{algorithm}

Algorithm \ref{opce_cash} provides a detailed description of the proposed OPCE-CASH method. Three important metrics are considered in the proposed OPCE-CASH method using MOPSO: F1-score, confidence, and model execution time. 

F1-score is a comprehensive classification performance metric that is commonly used in intrusion detection applications. As the F1-score represents the harmonic mean of precision and recall \cite{thesis}, optimizing it can result in an effective IDS that minimizes false positives and false negatives. Other metrics, like accuracy, may have biased results due to class-imbalance issues in network security datasets.

The average confidence metric, or prediction probability, is the second metric used in the proposed method. In ML, confidence refers to the likelihood or certainty of a prediction \cite{ml1}. For example, a confidence value of 80\% indicates that the ML model is 80\% certain or confident that this prediction is correct. Many ML algorithms, such as tree-based algorithms and neural network models, output confidence values with predictions \cite{eaai}. In the proposed OPCE-CASH approach, confidence values are optimized to ensure that the learning process is based on robust statistical evidence rather than arbitrary labels, thereby improving the reliability and robustness of IDS. 

Model execution time is the last metric considered in the model learning process, which involves both the model training and testing time \cite{mth}. By optimizing model execution time, MOO reduces the complexity of models, which directly indicates model learning efficiency. Model execution time is especially important for large network datasets and AutoML processes that often require complex models and multiple model trainings. Specifically, optimizing model training time can reduce resource and energy consumption on IoT servers, while optimizing model testing or inference time ensures that cyber threats are detected promptly to minimize damage. 

In the OPCE-CASH process illustrated in Algorithm \ref{opce_cash}, each MOPSO particle encodes both a selected ML model (\textit{e.g.}, XGBoost or LightGBM) and a vector of associated hyperparameter values. The position vector of the particle represents a specific model–hyperparameter configuration, and the velocity vector governs the rate and direction of exploration in the hyperparameter space. The fitness of each particle is evaluated using a three-dimensional objective vector: (1) F1-score (model effectiveness), (2) average prediction confidence (model reliability), and (3) inference latency (model efficiency). F1-score and confidence are naturally bounded in the range [0,1]. To match this scale, latency is normalized using a sigmoid transformation. This ensures that all three objectives lie within a comparable numeric range and can be jointly optimized without dominance bias. As this work aims to build a generic and reusable AutoML framework, we assign equal weights to all three normalized objectives by default. However, users may reweight these components depending on their application-specific constraints (\textit{e.g.}, low-latency devices may prioritize efficiency over accuracy). The MOPSO engine dynamically updates particle velocities and positions based on both local (personal best) and global (non-dominated archive) guidance, facilitating effective convergence toward well-balanced IDS solutions that offer trade-offs between model effectiveness, speed, and reliability.

In summary, the optimization of multiple objectives, the F1-score that indicates ML model performance, the confidence value that indicates model reliability, and the execution time that represents model complexity or efficiency, provides an innovative framework for developing IDSs that are both effective and efficient for resource-constrained systems like IoT. To the best of our knowledge, this is the first IDS optimization module that simultaneously optimizes F1-score, average prediction confidence, and inference latency within a CASH process, with the goal of enabling reliable deployment on resource-constrained systems. Moreover, the automation of all ML procedures in the overall MOO-AutoML IDS framework, including AutoDP, AutoFE, and automated model optimization, empowers autonomous cybersecurity solution development for modern and future networks with high automation demands \cite{mirna1}.

\section{Performance Evaluation}
\subsection{Experimental Setup and Dataset Description}

\begin{table}[!t]
\caption{Composition of The CICIDS2017 Dataset.}
\centering
\setlength\extrarowheight{1pt}
\scalebox{0.80}{
\begin{tabular}{|>{\centering\arraybackslash}p{5.5em}|
                >{\centering\arraybackslash}p{4em}|
                >{\centering\arraybackslash}p{5.3em}|
                >{\centering\arraybackslash}p{4em}|
                >{\centering\arraybackslash}p{5.5em}|
                >{\centering\arraybackslash}p{4em}|}
\hline
\textbf{Class Label (Attack Type or Normal)} & \textbf{Sample Counts} & \textbf{Class Distribution (\%)} & \textbf{Original Training Sample Counts} & \textbf{Training Sample Counts after AutoDP} & \textbf{Test Sample Counts} \\
\hline
Normal & 18,225 & 68.004 & 14,569 & 14,569 & 3,656 \\
\hline
DoS & 3,042 & 11.351 & 2,430 & 2,430 & 612 \\
\hline
Web-Attack & 2,180 & 8.134 & 1,728 & 1,728 & 452 \\
\hline
Botnet & 1,966 & 7.336 & 1,579 & 1,579 & 387 \\
\hline
Port Scan & 1,255 & 4.683 & 1,024 & 1,531 & 231 \\
\hline
Brute Force & 96 & 0.358 & 82 & 1,531 & 14 \\
\hline
Infiltration & 36 & 0.134 & 28 & 1,531 & 8 \\
\hline
\end{tabular}
}
\label{cicids2017_composition}
\end{table}

\begin{table}[!t]
\caption{Composition of The IoTID20 Dataset. }
\centering
\setlength\extrarowheight{1pt}
\scalebox{0.80}{
\begin{tabular}{|>{\centering\arraybackslash}p{5.5em}|
                >{\centering\arraybackslash}p{4em}|
                >{\centering\arraybackslash}p{5.3em}|
                >{\centering\arraybackslash}p{4em}|
                >{\centering\arraybackslash}p{5.5em}|
                >{\centering\arraybackslash}p{4em}|}
\hline
\textbf{Class Label (Attack Type or Normal)} & \textbf{Sample Counts} & \textbf{Class Distribution (\%)} & \textbf{Original Training Sample Counts} & \textbf{Training Sample Counts after AutoDP} & \textbf{Test Sample Counts} \\
\hline
Normal & 2,030 & 6.488 & 1,629 & 2,503 & 401 \\
\hline
Mirai & 20,780 & 66.413 & 16,637 & 16,637 & 4,143 \\
\hline
Scan & 3,720 & 11.889 & 2,961 & 2,961 & 759 \\
\hline
DoS & 2,994 & 9.569 & 2,389 & 2,503 & 605 \\
\hline
MITM & 1,765 & 5.641 & 1,415 & 2,503 & 350 \\
\hline
\end{tabular}
}
\label{iotid20_composition}
\end{table}

To develop the proposed MOO-AutoML IDS, we implemented the models by extending the Scikit-learn \cite{sklearn}, Xgboost \cite{xgboost}, Lightgbm \cite{lightgbm}, and Imbalanced-Learn \cite{imlearn} libraries within the Python programming environment. A Dell Precision 3630 based on an i7-8700 processor with 16 GB of Random Access Memory (RAM) was used for the experiments as an IoT server machine for cloud/edge data processing.

The proposed MOO-AutoML IDS framework is evaluated on two public benchmark network security datasets: CICIDS2017 \cite{cicdata} and IoTID20 \cite{iotid}. The CICIDS2017 \cite{cicdata} dataset from the Canadian Institute for Cybersecurity (CIC) is chosen for the IDS development due to its realistic and diverse nature. It provides an authentic environment for evaluating IDS by capturing various benign and malicious network behaviors. A wide range of cyber threats are covered by the dataset, including Denial of Service (DoS), botnets, brute force, port scanning, infiltration, and web attacks, as shown in Table \ref{cicids2017_composition}. The variety of these attacks makes it ideal for academic research and practical applications in security, as it can be applied to a wide range of IDS scenarios. 

The IoTID20 dataset \cite{iotid} represents a significant advancement in cybersecurity datasets for evaluating IDSs in IoT environments. By capturing network traffic from a smart home environment with interconnected IoT devices, this dataset addresses the increasing demand for realistic IoT cybersecurity datasets. A wide range of attack scenarios is available with IoTID20, which includes DoS, Man-in-the-Middle (MITM), scanning, and Mirai botnet variants, as shown in Table \ref{iotid20_composition}. In this dataset, 83 network-based features are extracted using the CICFlowMeter tool, and the attack type labels are used for the proposed IDS development. IoTID20 is an important resource for developing ML-based IDSs, especially for IoT security research, because of its well-structured features and real-world attack scenarios. As this research focuses on IDS development for resource-constrained systems like IoT, IoTID20 is a well-suited dataset for evaluating the proposed model.

Given the focus of this work on deployment in resource-constrained environments (\textit{e.g.}, IoT systems), representative subsets of CICIDS2017 and IoTID20 were drawn from the IDS-ML GitHub repository \cite{tii} \cite{idsml} and used for model evaluation. Stratified random sampling at the class level was performed to ensure statistical representativeness while reducing computational load. Stratified random sampling involves proportionally and randomly selecting samples from each class to form a representative subset that contains sufficient samples from each attack type and normal state. This subset-based design preserves diversity and attack patterns while aligning with real-world storage and processing constraints of edge devices. A representative subset of IoTID20 containing 31,289 samples, and a sampled subset of CICIDS2017 containing 26,800 samples, were generated and utilized. During the training and optimization process of the proposed MOO and AutoML-based IDS model, five-fold cross-validation was performed on the training set to obtain optimized ML models, and an 80\%/20\% hold-out validation split was applied in the testing process to evaluate the model on an unseen test set. 
Tables \ref{cicids2017_composition} and \ref{iotid20_composition} provide the composition of the CICIDS2017 and IoTID20 datasets used in this study. They include the attack types, total sample counts, class distributions, training sample counts before and after applying the proposed AutoDP method for data balancing, and test sample counts.

For comprehensively evaluating the proposed model’s effectiveness, four performance measures are used, including accuracy, precision, recall, and F1-scores, as network traffic data are often highly imbalanced and contain only a small proportion of attack samples \cite{icc}.  Additionally, since the proposed IDS emphasizes the balance between model effectiveness and efficiency, the execution time of the proposed model, including the total training time on the training set and the inference time per sample on the test set, is utilized to evaluate the proposed model's efficiency. The final ML model size is also measured to assess its storage footprint and suitability for deployment on resource-constrained devices. As resource-constrained network devices (\textit{e.g.}, IoT devices) often have limited memory, it is crucial to evaluate the practical applicability of IDS models \cite{iotop1} - \cite{iotop3}.
Furthermore, as model confidence is also optimized by the MOPSO method in the OPCE-CASH process to ensure model reliability, two important confidence metrics, average prediction probability and Expected Calibration Error (ECE), are used to quantify model confidence. Average prediction probability indicates the mean probability assigned to the true class across all test samples, while ECE quantifies the gap between predicted confidence and actual accuracy. Higher average prediction probability and lower ECE indicate more reliable models.

Lastly, for the purpose of model performance comparison, we have reproduced the methods in the literature \cite{ccs} \cite{ml2} \cite{ml3} \cite{oeids} \cite{automlid} \cite{icc} \cite{psolstm} \cite{svm} \cite{xgboost} \cite{lightgbm} on the selected datasets to comprehensively compare the proposed framework with the state-of-the-art IDSs using all the performance, efficiency, and confidence metrics. 

\subsection{Experimental Results and Analysis}

\begin{table*}[!t]
\caption{Performance Evaluation of the Proposed Models (including ablation studies) and State-of-the-Art Methods on the CICIDS2017 Dataset.}
\centering
\renewcommand{\arraystretch}{1.2}
\setlength\extrarowheight{1pt}
\scalebox{0.745}{
\begin{tabular}{|>{\centering\arraybackslash}p{7.5em}|
                >{\centering\arraybackslash}p{15em}|
                >{\centering\arraybackslash}p{4.2em}|
                >{\centering\arraybackslash}p{4.2em}|
                >{\centering\arraybackslash}p{3.8em}|
                >{\centering\arraybackslash}p{3.8em}|
                >{\centering\arraybackslash}p{4.5em}|
                >{\centering\arraybackslash}p{7.5em}|
                >{\centering\arraybackslash}p{5em}|
                >{\centering\arraybackslash}p{7em}|
                >{\centering\arraybackslash}p{4em}|}
\hline
\textbf{Category} & \textbf{Method} & \textbf{Accuracy (\%)} & \textbf{Precision (\%)} & \textbf{Recall (\%)} & \textbf{F1 (\%)} & \textbf{Training Time (s)} & \textbf{Avg Test Time Per Sample (ms)} & \textbf{Model Size (MB)} & \textbf{Avg Confidence (\%)} & \textbf{ECE (\%)} \\ \hline

\multirow{8}{=}{Literature \\ (For Comparison)} 
& GA\text{-}RF \cite{ml2} & 99.235 & 99.240 & 99.235 & 99.224 & 2.27  & 0.0097 & 5.50 & 97.65 & 0.28 \\ \cline{2-11}
& AutoEncoder\text{-}RF \cite{ml3} & 98.377 & 98.354 & 98.377 & 98.357 & 28.62 & 0.0102 & 10.33 & 97.99 & 0.17 \\ \cline{2-11}
& PSO\text{-}CNN \cite{icc} & 91.959 & 92.079 & 91.959 & 91.552 & 66.93 & 0.1785 & 1.71 & 95.58 & 0.68 \\ \cline{2-11}
& PSO\text{-}LSTM \cite{psolstm} & 91.045 & 91.177 & 91.045 & 90.942 & 1623.62 & 4.6900 & 1.44 & 89.16 & 1.98 \\ \cline{2-11}
& MOPSO\text{-}SVM \cite{svm} & 91.007 & 90.755 & 91.007 & 90.646 & 8.59 & 0.6064 & 3.75 & 93.77 & 0.34 \\ \cline{2-11}
& OE\text{-}IDS \cite{oeids} & 99.254 & 99.257 & 99.254 & 99.240 & 115.74 & 0.0377 & 32.83 & 96.25 & 0.68 \\ \cline{2-11}
& AutoML\text{-}ID \cite{automlid} & 99.515 & 99.508 & 99.515 & 99.511 & 72.73 & 0.0078 & 1.52 & 98.80 & 0.10 \\ \cline{2-11}
& Our Previous AutoML Model \cite{ccs} & 99.720 & 99.721 & 99.720 & 99.708 & 41.56 & 0.0158 & 7.71 & 99.69 & 0.04 \\ \hline

\multirow{14}{=}{Proposed Framework \\ (with Ablation Studies)}
& XGBoost \cite{xgboost} & 99.720 & 99.721 & 99.720 & 99.708 & 5.77 & 0.0026 & 0.91 & 99.81 & 0.05 \\ \cline{2-11}
& LightGBM \cite{lightgbm} & 99.590 & 99.594 & 99.590 & 99.586 & 1.27 & 0.0058 & 2.23 & 99.90 & 0.02 \\ \cline{2-11}
& AutoDP + XGBoost & 99.739 & 99.739 & 99.739 & 99.736 & 6.58 & 0.0022 & 0.93 & 99.82 & 0.05 \\ \cline{2-11}
& AutoDP + LightGBM & 99.739 & 99.740 & 99.739 & 99.736 & 2.00 & 0.0070 & 2.38 & 99.89 & 0.04 \\ \cline{2-11}
& AutoDP + AutoFE + XGBoost & 99.720 & 99.721 & 99.720 & 99.714 & 3.01 & 0.0019 & 1.34 & 99.80 & 0.03 \\ \cline{2-11}
& AutoDP + AutoFE + LightGBM & 99.720 & 99.721 & 99.720 & 99.714 & 1.00 & 0.0053 & 2.22 & 99.78 & 0.07 \\ \cline{2-11}
& AutoDP + CASH + XGBoost & 99.720 & 99.721 & 99.720 & 99.718 & 5.31 & 0.0020 & 1.02 & 99.77 & 0.06 \\ \cline{2-11}
& AutoDP + CASH + LightGBM & 99.720 & 99.721 & 99.720 & 99.718 & 1.15 & 0.0030 & 0.63 & 99.90 & 0.02 \\ \cline{2-11}
& \textbf{Full Pipeline (AutoDP + AutoFE + CASH) + XGBoost} & \textbf{99.776} & \textbf{99.777} & \textbf{99.776} & \textbf{99.773} & 2.82 & \textbf{0.0016} & 1.12 & 99.83 & 0.02 \\ \cline{2-11}
& \textbf{Full Pipeline (AutoDP + AutoFE + CASH) + LightGBM} & 99.757 & 99.758 & 99.757 & 99.751 & \textbf{0.33} & 0.0031 & \textbf{0.42} & \textbf{99.91} & \textbf{0.01} \\ \hline
\end{tabular}
}
\label{cicids2017_table}
\end{table*}

\begin{table*}[!t]
\caption{Performance Evaluation of the Proposed Models (including ablation studies) and State-of-the-Art Methods on the IoTID20 Dataset.}
\centering
\renewcommand{\arraystretch}{1.2}
\setlength\extrarowheight{1pt}
\scalebox{0.745}{
\begin{tabular}{|>{\centering\arraybackslash}p{7.5em}|
                >{\centering\arraybackslash}p{15em}|
                >{\centering\arraybackslash}p{4.2em}|
                >{\centering\arraybackslash}p{4.2em}|
                >{\centering\arraybackslash}p{3.8em}|
                >{\centering\arraybackslash}p{3.8em}|
                >{\centering\arraybackslash}p{4.5em}|
                >{\centering\arraybackslash}p{7.5em}|
                >{\centering\arraybackslash}p{5em}|
                >{\centering\arraybackslash}p{7em}|
                >{\centering\arraybackslash}p{4em}|}
\hline
\textbf{Category} & \textbf{Method} & \textbf{Accuracy (\%)} & \textbf{Precision (\%)} & \textbf{Recall (\%)} & \textbf{F1 (\%)} & \textbf{Training Time (s)} & \textbf{Avg Test Time Per Sample (ms)} & \textbf{Model Size (MB)} & \textbf{Avg Confidence (\%)} & \textbf{ECE (\%)} \\ \hline

\multirow{8}{=}{Literature \\ (For Comparison)} 
& GA\text{-}RF \cite{ml2} & 98.114 & 98.114 & 98.114 & 98.082 & 2.61 & 0.0131 & 14.56 & 96.86 & 0.38 \\ \cline{2-11}
& AutoEncoder\text{-}RF \cite{ml3} & 96.868 & 96.853 & 96.868 & 96.851 & 44.82 & 0.0118 & 16.91 & 96.20 & 0.30 \\ \cline{2-11}
& PSO\text{-}CNN \cite{icc} & 78.108 & 77.408 & 78.108 & 73.613 & 81.96 & 0.1840 & 1.71 & 90.15 & 7.67 \\ \cline{2-11}
& PSO\text{-}LSTM \cite{psolstm} & 83.557 & 88.485 & 83.557 & 84.135 & 4,396.43 & 3.9730 & 1.60 & 82.92 & 4.52 \\ \cline{2-11}
& MOPSO\text{-}SVM \cite{svm} & 85.427 & 88.442 & 85.427 & 85.928 & 22.00 & 1.0971 & 5.49 & 84.87 & 2.39 \\ \cline{2-11}
& OE\text{-}IDS \cite{oeids} & 97.507 & 97.536 & 97.507 & 97.444 & 69.81 & 0.0453 & 69.65 & 93.50 & 1.28 \\ \cline{2-11}
& AutoML\text{-}ID \cite{automlid} & 98.865 & 98.861 & 98.865 & 98.859 & 49.31 & 0.0057 & 0.86 & 98.94 & 0.09 \\ \cline{2-11}
& Our Previous AutoML Model \cite{ccs} & 99.169 & 99.169 & 99.169 & 99.166 & 33.62 & 0.0133 & 5.73 & 99.08 & 0.07 \\ \hline

\multirow{14}{=}{Proposed Framework \\ (with Ablation Studies)}
& XGBoost \cite{xgboost} & 99.169 & 99.165 & 99.169 & 99.165 & 4.75 & 0.0020 & 1.09 & 99.46 & 0.03 \\ \cline{2-11}
& LightGBM \cite{lightgbm} & 98.673 & 98.671 & 98.671 & 98.660 & 1.19 & 0.0042 & 1.65 & 99.38 & 0.10 \\ \cline{2-11}
& AutoDP + XGBoost & 99.249 & 99.246 & 99.249 & 99.244 & 7.76 & 0.0038 & 1.17 & 99.40 & 0.07 \\ \cline{2-11}
& AutoDP + LightGBM & 98.897 & 98.893 & 98.897 & 98.889 & 1.19 & 0.0053 & 1.65 & 99.38 & 0.10 \\ \cline{2-11}
& AutoDP + AutoFE + XGBoost & 99.249 & 99.246 & 99.249 & 99.244 & 2.26 & 0.0018 & 1.14 & 99.26 & 0.13 \\ \cline{2-11}
& AutoDP + AutoFE + LightGBM & 98.897 & 98.893 & 98.897 & 98.889 & 0.83 & 0.0038 & 1.62 & 99.15 & 0.10 \\ \cline{2-11}
& AutoDP + CASH + XGBoost & 99.201 & 99.197 & 99.201 & 99.195 & 6.04 & 0.0022 & 1.78 & 99.56 & 0.06 \\ \cline{2-11}
& AutoDP + CASH + LightGBM & 98.801 & 98.798 & 98.801 & 98.790 & 0.58 & 0.0036 & 0.32 & 98.50 & 0.14 \\ \cline{2-11}
& \textbf{Full Pipeline (AutoDP + AutoFE + CASH) + XGBoost} & \textbf{99.313} & \textbf{99.321} & \textbf{99.313} & \textbf{99.309} & 1.47 & \textbf{0.0015} & 0.53 & \textbf{99.58} & \textbf{0.02} \\ \cline{2-11}
& \textbf{Full Pipeline (AutoDP + AutoFE + CASH) + LightGBM} & 98.913 & 98.912 & 98.913 & 98.903 & \textbf{0.46} & 0.0030 & \textbf{0.29} & 99.46 & 0.03 \\ \hline
\end{tabular}
}
\label{iotid20_table}
\end{table*}

\subsubsection{Comparison of Model Effectiveness with State-of-the-Art Methods}

The performance comparison between the proposed MOO-AutoML IDS and representative state-of-the-art methods from the literature \cite{ccs} \cite{ml2} \cite{ml3} \cite{oeids} \cite{automlid} \cite{icc} \cite{psolstm} \cite{svm} \cite{xgboost} \cite{lightgbm} on the CICIDS2017 and IoTID20 datasets is presented in detail in Tables \ref{cicids2017_table} and \ref{iotid20_table}, respectively. The selected comparison methods cover three important categories of techniques, including deep learning, feature selection, and optimization or AutoML-based IDS models. Among the state-of-the-art methods used for comparison, AutoEncoder-RF \cite{ml3}, PSO-CNN \cite{icc}, and PSO-LSTM \cite{psolstm} are based on deep learning models employing autoencoders, CNNs, and LSTMs, respectively. GA-RF \cite{ml2}, AutoEncoder-RF \cite{ml3}, and MOPSO-SVM \cite{svm} introduce novel feature selection methods based on GA, autoencoder, and MOPSO, respectively, to effectively identify and select important features. Our previous AutoML model \cite{ccs}, PSO-CNN \cite{icc}, PSO-LSTM \cite{psolstm}, OE-IDS \cite{oeids}, and AutoML-ID \cite{automlid} are optimized models that use PSO or Bayesian optimization for model tuning. This broad selection ensures a comprehensive and fair evaluation across a wide range of methodological paradigms.

As shown in Table \ref{cicids2017_table}, on the CICIDS2017 dataset, the proposed MOO-AutoML IDS (\textit{i.e.}, Full Pipeline + XGBoost/LightGBM), which comprises AutoDP, AutoFE, and CASH components, outperforms all traditional and state-of-the-art models on the CICIDS2017 dataset. The optimized XGBoost achieves the highest accuracy (99.776\%), precision (99.777\%), recall (99.776\%), and F1-score (99.773\%). Based on F1-score, the comprehensive classification metric, the proposed optimized XGBoost model outperforms all baseline models in terms of F1-score, including optimized deep learning models (\textit{e.g.}, PSO-CNN \cite{icc} (91.522\%), PSO-LSTM \cite{psolstm} (90.942\%)), advanced feature selection-based IDS models (\textit{i.e.}, GA-RF \cite{ml2} (99.224\%), AutoEncoder-RF \cite{ml3}(98.357\%), and MOPSO-SVM \cite{svm})(90.646\%)), and optimized \& automated ML models (\textit{e.g.}, OE-IDS \cite{oeids} (99.240\%), AutoML-ID \cite{automlid} (99.511\%), and our previous AutoML model \cite{ccs} (99.708\%)).

Similarly, Table \ref{iotid20_table} shows the comparison between the proposed IDS model and the state-of-the-art methods in the same literature \cite{ccs} \cite{ml2} \cite{ml3} \cite{oeids} \cite{automlid} \cite{icc} \cite{psolstm} \cite{svm} \cite{xgboost} \cite{lightgbm} on the IoTID20 dataset. According to Table \ref{iotid20_table}, the proposed optimized XGBoost model (\textit{i.e.}, Full Pipeline) achieves the best overall results, with an accuracy of 99.313\%, precision of 99.321\%, recall of 99.313\%, and an F1-score of 99.309\%. The proposed model outperforms all state-of-the-art methods in the literature. This includes optimized deep learning models (\textit{e.g.}, PSO-CNN \cite{icc} (73.613\%), PSO-LSTM \cite{psolstm} (84.135\%)); advanced feature selection-based IDS models (\textit{i.e.}, GA-RF \cite{ml2} (98.082\%), AutoEncoder-RF \cite{ml3} (96.851\%), MOPSO-SVM \cite{svm} (85.928\%)); and optimized \& automated ML models (\textit{e.g.}, OE-IDS \cite{oeids} (97.444\%), AutoML-ID \cite{automlid} (98.859\%), our previous AutoML model \cite{ccs} (99.166\%)). Furthermore, the optimized LightGBM model has a high F1-score (98.903\%), surpassing the original LightGBM model (98.660\%) and all other state-of-the-art methods, except for our previous AutoML model (99.166\%). The performance improvement of the proposed model on IoTID20 highlights the robustness of the proposed framework to IoT-specific traffic data.

\subsubsection{Comparison of Model Efficiency with State-of-the-Art Methods}

Efficient execution time and compact model sizes are critical factors in the practical deployment of IDSs, particularly in resource-constrained IoT environments where computational resources, memory, and network bandwidth are limited. The proposed MOO-AutoML IDS framework is explicitly designed to optimize both detection effectiveness and computational efficiency, balancing training time, inference latency, and model size. These metrics must be interpreted based on the intended deployment scenario, whether frequent local retraining on edge devices or centralized training with model distribution to the edge.

Training time measures the computational cost required to train an IDS model on network traffic data, while model size measures the deployability of an IDS model on devices with limited memory and storage. This consideration is especially crucial for IoT-edge applications, where ML models may be trained locally. As observed in Tables \ref{cicids2017_table} and \ref{iotid20_table}, the optimized LightGBM model (Full Pipeline + LightGBM) achieves the shortest training time among all compared methods (0.33 s on CICIDS2017, 0.46 s on IoTID20) and smallest model sizes (0.42 megabytes (MB) on CICIDS2017, 0.29 MB on IoTID20), making it well-suited for edge-based IoT applications where models must be trained locally on edge devices with limited computational resources. Additionally, the optimized XGBoost with a short training time of 2.82 s on CICIDS2017 and 2.47 s on IoTID20 and a low model size of 1.12 MB on CICIDS2017 and 0.53 MB on IoTID20 achieves a favorable trade-off between model effectiveness and training efficiency, making it ideal for cloud-based IDS deployments, where models can be trained on the cloud with greater computational resources and then deployed to multiple edge devices for inference.

Comparatively, deep learning models such as PSO-LSTM \cite{psolstm} (1623.62 s training time on CICIDS2017, 4396.43 s on IoTID20) and PSO-CNN \cite{icc} (66.93 s training time on CICIDS2017, 81.96 s on IoTID20) require a significantly longer training time due to the high computational complexity of their complex DL architectures, which requires extensive parameter tuning and prolonged training iterations. Similarly, models such as OE-IDS \cite{oeids} (115.74 s on CICIDS2017, 69.81 s on IoTID20) and AutoML-ID \cite{automlid} (72.73 s on CICIDS2017, 49.31 s on IoTID20) demand longer training times, making them less suitable for efficient model training in edge IoT devices. Moreover, the model sizes of OE-IDS \cite{oeids} (32.83 MB on CICIDS2017 and 69.65 MB on IoTID20) and our previous AutoML model \cite{ccs} (7.71 MB on CICIDS2017 and 5.73 MB on IoTID20) are relatively larger because they are built on ensemble models that combine multiple base ML models.

Inference time is the amount of time it takes to classify a network traffic sample after the model has been trained. For real-time cybersecurity applications, this feature is essential, particularly in cloud-edge architectures where cloud models are trained and deployed on edge devices. As shown in Tables \ref{cicids2017_table} and \ref{iotid20_table}, the optimized XGBoost model achieves the lowest inference time (0.0016 ms on CICIDS2017, 0.0015 ms on IoTID20), significantly outperforming traditional models such as GA-RF \cite{ml2} (0.0097 ms on CICIDS2017, 0.0131 ms on IoTID20), AutoML-ID \cite{automlid} (0.0078 ms on CICIDS2017, 0.0057 ms on IoTID20), and PSO-CNN \cite{icc} (0.1785 ms on CICIDS2017, 0.1840 ms on IoTID20). Additionally, models such as PSO-LSTM \cite{psolstm} (4.6900 ms on CICIDS2017, 3.9730 ms on IoTID20) and MOPSO-SVM \cite{svm} (0.6064 ms on CICIDS2017, 1.0971 ms on IoTID20) exhibit significantly longer inference times, which make them unsuitable for real-time cybersecurity applications requiring immediate threat detection and response.

The high efficiency and small model size of the proposed optimized XGBoost and LightGBM are primarily due to the proposed OIP-AutoFE method that reduces the dimensionality of the dataset while preserving the most important features, as well as the proposed MOPSO-Based OPCE-CASH method, which identifies ML model hyperparameter configurations that maximize the F1 scores and confidence while minimizing computational cost. This ensures that models can be trained efficiently without excessive resource consumption.

\subsubsection{Comparison of Model Reliability (Confidence) with State-of-the-Art Methods}

Confidence and calibration are critical for the operational reliability of IDSs, as they determine how trustworthy the model’s predicted probabilities are in reflecting the true likelihood of network events. High average confidence is desirable when predictions are correct, but without proper calibration, overconfident and incorrect predictions can lead to dangerous misjudgments, such as ignoring actual threats or generating excessive false alarms. The Expected Calibration Error (ECE) is therefore an essential complement to average confidence, quantifying the degree of misalignment between predicted confidence and empirical accuracy.

As shown in Table~\ref{cicids2017_table} for the CICIDS2017 dataset, the proposed model (Full Pipeline + LightGBM) achieves the highest average confidence at 99.91\% with an ECE of only 0.01\%, representing the most reliable and well-calibrated predictions among all methods. This ECE value is 10$\times$ lower than AutoML-ID \cite{automlid} (0.10\%), 4$\times$ lower than our previous AutoML \cite{ccs} (0.04\%), and substantially better than GA-RF \cite{ml2} (0.28\%) and AutoEncoder-RF \cite{ml3} (0.17\%). Deep learning models exhibit both lower confidence and significantly worse calibration: PSO-CNN \cite{icc} reports an average confidence of 95.58\% with an ECE of 0.68\%, while PSO-LSTM \cite{psolstm} is poorly calibrated (89.16\% confidence, 1.98\% ECE). These results indicate that, beyond achieving the highest classification performance, the proposed framework also produces the most reliable probability estimates for security decision-making, due to the proposed MOO-based OPCE-CASH component that optimizes model prediction confidence.

On the IoTID20 dataset, as shown in Table~\ref{iotid20_table}, the proposed model (Full Pipeline + XGBoost) delivers the best overall balance of confidence and calibration, with an average confidence of 99.58\% and an ECE of only 0.02\%. This reduces calibration error by 4.5$\times$ compared to AutoML-ID (0.09\%), 3.5$\times$ compared to our previous AutoML (0.07\%), and by 19$\times$ compared to GA-RF (0.38\%). AutoEncoder-RF \cite{ml3} shows both lower confidence (96.20\%) and worse calibration (0.30\%), while deep learning methods again show severe miscalibration, with PSO-CNN \cite{icc} reaching an ECE of 7.67\% and PSO-LSTM \cite{psolstm} at 4.52\%, indicating a substantial gap between their predicted probabilities and actual correctness. Even among optimized AutoML baselines, none match the proposed pipeline’s combination of high confidence and minimal calibration error.

The improvement in calibration performance can be attributed to several design factors. First, the OPCE-CASH stage explicitly incorporates prediction confidence into the multi-objective optimization process, jointly optimizing for F1-score, average confidence, and inference latency. This ensures that selected model configurations are not only accurate but also probabilistically reliable. Second, the OIP-AutoFS step reduces feature redundancy and noise, which are known causes of overconfident misclassifications, leading to cleaner decision boundaries and better-aligned probability outputs. Third, the choice of tree-based models (\textit{i.e.}, XGBoost, LightGBM) as the base classifiers inherently supports better probability calibration when hyperparameters such as tree depth and learning rate are tuned appropriately.

\subsubsection{Model Comparison Results Analysis}

Overall, the proposed MOO-AutoML IDS achieves higher accuracy, precision, recall, F1-score, and average confidence, and lower training time, average test time per sample, model size, and ECE than all state-of-the-art methods in the literature \cite{ccs} \cite{ml2} \cite{ml3} \cite{oeids} \cite{automlid} \cite{icc} \cite{psolstm} \cite{svm} \cite{xgboost} \cite{lightgbm}. These consistent improvements across both the CICIDS2017 and IoTID20 datasets demonstrate that the proposed framework not only delivers superior detection capability but also offers clear advantages in efficiency and reliability, which are crucial for practical IDS deployment in real-world resource-constrained and real-time network environments.

First, the proposed MOO-AutoML framework outperforms deep learning–based IDS models, such as AutoEncoder-RF \cite{ml3}, PSO-CNN \cite{icc}, and PSO-LSTM \cite{psolstm}, primarily because it leverages GBDT-based ML algorithms (XGBoost and LightGBM) as base classifiers. Unlike deep learning models, which often require large amounts of balanced data and may overfit on highly imbalanced tabular network traffic, GBDT models inherently handle heterogeneous feature types, missing values, and non-linear feature interactions more effectively. Furthermore, they are less sensitive to extensive hyperparameter tuning and can achieve optimal performance with smaller, more interpretable model architectures, resulting in reduced model size and faster inference. The integration of AutoDP ensures that data preprocessing is fully aligned with model requirements, mitigating class imbalance and feature-scale disparities that often hinder deep learning approaches in IDS development.

\begin{figure}
     \centering
     \includegraphics[width=\columnwidth]{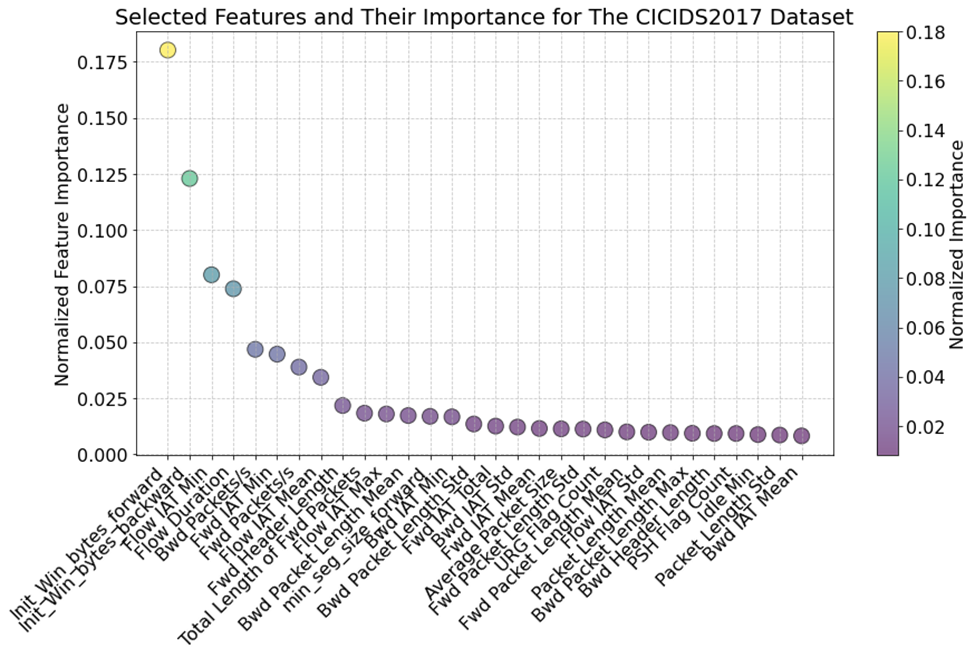}
     \caption{The selected features and their importance scores for the CICIDS2017 dataset.} 
     \label{fs_cic}
\end{figure}

\begin{figure}
     \centering
     \includegraphics[width=\columnwidth]{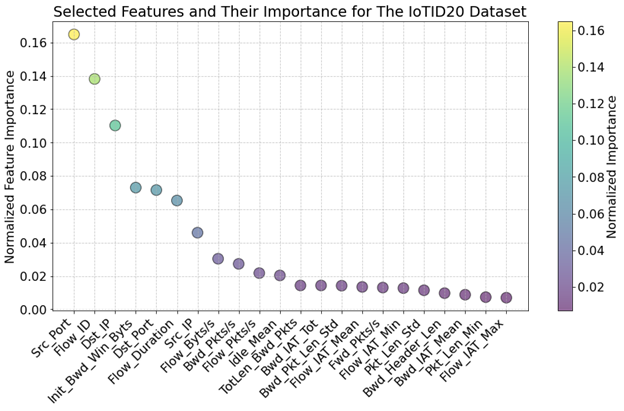}
     \caption{The selected features and their importance scores for the IoTID20 dataset.} 
     \label{fs_iot}
\end{figure}

Second, the proposed MOO-AutoML framework outperforms advanced feature selection methods such as GA-RF \cite{ml2}, AutoEncoder-RF \cite{ml3}, and MOPSO-SVM \cite{svm} because it incorporates the OIP-AutoFS mechanism, which is grounded in an IG–based evaluation of features but enhanced through a learner-aware optimization process. The use of IG in OIP-AutoFS offers a clear advantage over other feature selection methods by directly quantifying each feature’s contribution to reducing class uncertainty, which is particularly effective for high-dimensional, heterogeneous intrusion detection data. When integrated into a learner-aware optimization process, IG not only identifies the most discriminative features but also ensures that the selected subset is synergistic with the base classifier, avoiding the suboptimality often seen in filter- or wrapper-based FS methods used in prior works. Additionally, the proposed framework strikes a trade-off between accumulated feature importance and data dimensionality. The selected features and their importance scores for the two datasets are shown in Figs. \ref{fs_cic} and \ref{fs_iot}. This strategy minimizes redundant or noisy features, producing smaller, faster models without compromising detection accuracy, especially for minority attack classes.

Third, the framework improves upon existing AutoML-based and optimization-based IDS approaches, including our previous AutoML model \cite{ccs}, OE-IDS \cite{oeids}, AutoML-ID \cite{automlid}, and optimization-based deep learning models such as PSO-CNN \cite{icc} and PSO-LSTM \cite{psolstm}, through its novel OPCE-CASH mechanism. Unlike our prior work \cite{ccs}, which adopted a single-objective Bayesian optimization strategy focusing primarily on F1-score, the proposed MOO-AutoML model employs a multi-objective formulation that simultaneously optimizes detection effectiveness, prediction confidence, and computational efficiency. This approach ensures the selection of configurations that are not only accurate but also well-calibrated and lightweight, making them suitable for deployment in resource-constrained environments. The integration of OIP-AutoFS with OPCE-CASH further enables joint optimization of feature selection, model choice, and hyperparameters under deployment-aware constraints, producing models that are more accurate, faster, smaller, and better calibrated than those in our previous work \cite{ccs}, thereby supporting flexible deployment from centralized systems to IoT-edge devices.

\subsubsection{Ablation Studies and Component Impact}
To quantify the contribution of each component in the proposed MOO-AutoML IDS pipeline, we conducted a series of ablation experiments on both the CICIDS2017 and IoTID20 datasets. The detailed results in Tables~\ref{cicids2017_table} and \ref{iotid20_table} enable precise measurement of the performance gains in detection effectiveness (F1-score, accuracy, precision, recall), efficiency (training time, inference latency, model size), and reliability (average confidence, ECE).

Firstly, it is worth mentioning that the original XGBoost and LightGBM models achieve higher F1-scores than most other models in the literature \cite{ml2} \cite{ml3} \cite{oeids} \cite{automlid} \cite{icc} \cite{psolstm} \cite{svm} \cite{xgboost} \cite{lightgbm} on both datasets, which further supports their selection as base classifiers in the proposed IDS framework. Additionally, tree-based algorithms in the literature \cite{ml2} \cite{ml3} \cite{xgboost} \cite{lightgbm} outperform other ML models like SVM \cite{svm}, LSTM \cite{psolstm}, and CNN \cite{icc}, justifying their selection for the proposed method.

As shown in Table~\ref {cicids2017_table}, on CICIDS2017, starting from the original XGBoost with an F1-score of 99.708\%, adding AutoDP improves the F1 marginally to 99.736\%, while also improving confidence from 99.81\% to 99.82\%. Adding the AutoFE component (OIP-AutoFS) on top of AutoDP yields a lower F1 of 99.714\% but reduces the model training time from 6.58 s to 3.01 s and the average test time per sample from 0.0022 ms to 0.0019 ms due to the reduction in feature dimensionality. 
Incorporating OPCE-CASH into the pipeline (Full Pipeline + XGBoost) achieves 99.776\% F1 and a better balance of confidence and calibration (99.83\% / 0.02\% ECE), demonstrating the value of MOO and hyperparameter tuning. The Full Pipeline + LightGBM model achieves a slightly lower F1 (99.751\%) than Full Pipeline + XGBoost, but a shorter training time of 0.33 s, a smaller model size of 0.42 MB, higher confidence of 99.91\%, and a lower ECE of 0.01\%. This shows that the joint use of AutoDP, AutoFS, and OPCE-CASH produces cumulative benefits.
Moreover, the full features, no AutoFS baseline (AutoDP + CASH) confirms that AutoFS is responsible for a measurable reduction in model training time (from 5.31 s to 2.82 s for XGBoost) and a drop in inference time (from 0.0020 ms to 0.0016 ms), with a higher F1 (from 99.718\% to 99.773\%) and a higher average confidence (from 99.77\% to 99.83\%). 

Similar trends are observed on IoTID20 in Table \ref{iotid20_table}. Starting from the original XGBoost (F1 = 99.165\%), AutoDP increases the F1-score to 99.244\% while slightly increasing inference time from 0.0020 ms to 0.0038 ms due to the adoption of synthetic minority class samples. Adding AutoFS on top of AutoDP yields the same F1-score but significantly reduces training time (7.76 s to 2.26 s), average test time per sample (0.0020 ms to 0.0018 ms), and model size (1.17 MB to 1.14 MB). By adding OPCE-CASH, the Full Pipeline + XGBoost achieves the highest F1 (99.309\%), the fastest inference time (0.0015 ms), the highest confidence (99.58\%), and the best calibration (ECE = 0.02\%), while the Full Pipeline + LightGBM achieves a relatively high F1 of 98.903\%, the fastest training time (0.46 s), and the smallest model sizes (0.29 MB).
Additionally, the “full features, no AutoFE” baseline (AutoDP + CASH) confirms that AutoFS is responsible for model compactness and marginal ECE improvement. Without the proposed OIP-AutoFS component, the F1-score of the optimized XGBoost model drops from 99.309\% to 99.195\%, training time increases from 1.47 s to 6.04 s, model size increases from 0.53 MB to 1.78 MB, and average confidence drops from 99.58\% to 99.56\%.

Overall, the proposed MOO-AutoML IDS models outperform other state-of-the-art methods due to the following advantages of each AutoML pipeline component:
\begin{enumerate}
\item In the proposed AutoDP process, the integration of SMOTE and ADASYN for automated data balancing ensures that the model does not suffer from majority-class bias, thereby increasing recall and intrusion detection effectiveness.
\item In the proposed OIP-AutoFS method, only the most significant network traffic features are used to train the model, while irrelevant, redundant, and noisy features are eliminated to enhance intrusion detection accuracy and efficiency. 
\item The proposed OPCE-CASH algorithm optimizes ML models automatically by balancing F1-score, execution time, and model confidence, which reduces computational complexity while increasing intrusion detection effectiveness and reliability.
\end{enumerate}

In summary, the proposed optimized LightGBM is best suited for edge computing in IoT systems, where IDS models need to be trained locally on low-power edge devices. Its high F1-score and fast training time allow effective and efficient retrainings based on the continuous arrival of network data. On the other hand, the optimized XGBoost model is well-suited to cloud-based IoT applications, where models can be trained periodically with sufficient computing resources and then distributed for edge deployment, which makes efficient predictions. Moreover, while the proposed model assigns the same weights to each objective in the MOO process, the weights of different objectives in the proposed MOO-based OIP-AutoFS and OPCE-CASH components can be adjusted to fit specific detection rate, training/inference time, and memory constraints. This approach enables global threat intelligence sharing while optimizing resource usage. With the proposed MOO-AutoML framework, comprehensive optimized IDS frameworks are available that are suitable for a wide range of IoT applications.

\section{Conclusion}
The continuous evolution of cybersecurity threats, particularly in resource-constrained environments like IoT, necessitates intelligent, autonomous, and computationally efficient intrusion detection mechanisms. This paper introduces the MOO-AutoML IDS, a novel intrusion detection framework that integrates MOO and AutoML techniques for IoT systems and modern networks. By integrating innovative automated data pre-processing, optimized feature selection, and model selection and optimization, the proposed approach effectively balances detection accuracy, computational efficiency, and model reliability. Experimental results on CICIDS2017 and IoTID20 datasets demonstrated that the proposed IDS models outperform state-of-the-art IDS models, achieving higher F1-scores and confidence while reducing training and inference overhead. The results also highlight key deployment advantages: the optimized LightGBM’s lower training time makes it ideal for edge computing in IoT environments, while the optimized XGBoost’s minimal inference time is better suited for cloud-edge architectures requiring real-time threat detection. Future research will explore continual learning and model adaptation techniques to enhance the framework’s ability to respond to evolving cyber threats in 5G and next-generation networks. Additionally, future work will explore constraint-based formulations (\textit{e.g.}, enforcing latency below a predefined threshold based on specific device configurations) using feasibility-constrained PSO or hybrid MOO techniques.



\begin{IEEEbiography}[{\includegraphics[width=1in,height=1.25in,clip,keepaspectratio]{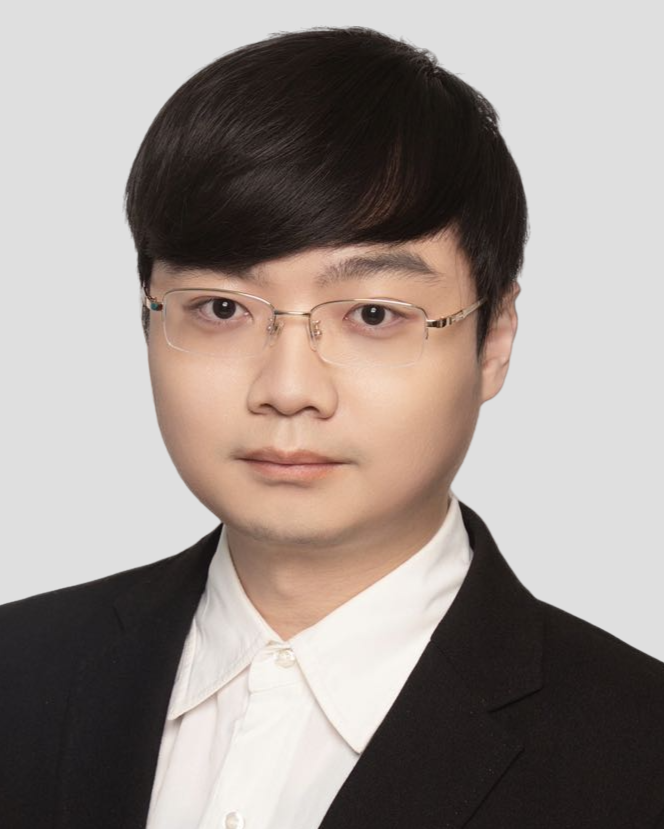}}]{Li Yang}\, (Member, IEEE) is currently an Assistant Professor in the Faculty of Business and Information Technology at Ontario Tech University, and an Adjunct Research Professor in the Department of Electrical and Computer Engineering at Western University. He received his Ph.D. in Electrical and Computer Engineering from Western University in 2022. He is currently an Associate Editor of IEEE Transactions on Industrial Informatics (TII) and IEEE Transactions on Network and Service Management (TNSM). He was the vice chair of the IEEE Computer Society, London Section, Canada, from 2022 to 2023. He was also on the technical program committee for IEEE GlobeCom from 2023 to 2025, the workshop chair for SMC-IoT 2023, and the technical session chair for IEEE CCECE 2020. His paper and code publications have received thousands of Google Scholar citations and GitHub stars. His research interests include cybersecurity, machine learning, deep learning, AutoML, model optimization, network data analytics, Internet of Things (IoT), intrusion detection, anomaly detection, concept drift, continual learning, and adversarial machine learning. Li Yang is also included in Stanford University/Elsevier's List of the World's Top 2\% Scientists. He was ranked among the world's Top 0.5\% of researchers in 'Networking \& Telecommunications' in 2024 and 2025.
\end{IEEEbiography}

\begin{IEEEbiography}[{\includegraphics[width=1in,height=1.25in,clip,keepaspectratio]{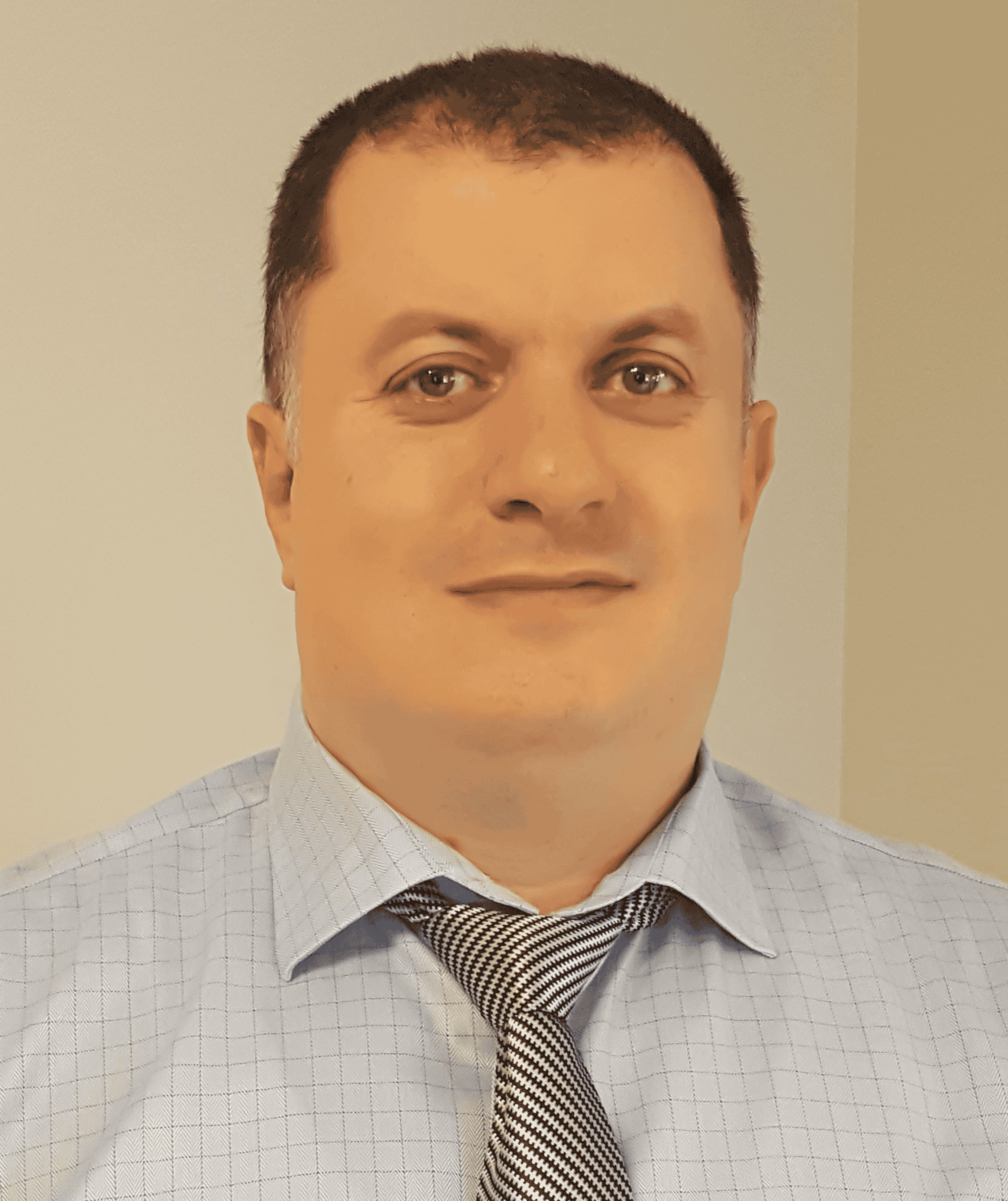}}] {Abdallah Shami}\, (Fellow, IEEE)  is currently a Professor and Chair of the Department of Electrical and Computer Engineering, Western University, London, ON, Canada, where he is also the Director of the Optimized Computing and Communications Laboratory. Dr. Shami has chaired key symposia for the IEEE GLOBECOM, IEEE International Conference on Communications, and IEEE International Conference on Computing, Networking and Communications. He was the elected Chair for the IEEE Communications Society Technical Committee on Communications Software and the IEEE London Ontario Section Chair. He is currently an Associate Editor of the IEEE Transactions on Information Forensics and Security, IEEE Transactions on Network and Service Management, and IEEE Communications Surveys and Tutorials journals. Dr. Shami is a Fellow of IEEE, a Fellow of the Canadian Academy of Engineering (CAE), and a Fellow of the Engineering Institute of Canada (EIC). 
\end{IEEEbiography}

\vfill\pagebreak

\end{document}